\documentclass[%
 reprint,
groupedaddress,
showpacs,
 amsmath,amssymb,
 aps,
prb,
]{revtex4-1}

\usepackage{graphicx}
\usepackage{dcolumn}
\usepackage{bm}

\begin{document}

\newcommand{\be}{\begin{equation}}
\newcommand{\ee}[1]{\label{#1}\end{equation}}
\newcommand{\bem}{\begin{eqnarray}}
\newcommand{\eem}[1]{\label{#1}\end{eqnarray}}
\newcommand{\eq}[1]{Eq.~(\ref{#1})}
\newcommand{\Eq}[1]{Equation~(\ref{#1})}
\newcommand{\vp}[2]{[\mathbf{#1} \times \mathbf{#2}]}


\title{Transverse force on a vortex and vortex mass: \\ effects of free bulk  and vortex-core bound quasiparticles}

\author{E.  B. Sonin}
 \affiliation{Racah Institute of Physics, Hebrew University of
Jerusalem, Givat Ram, Jerusalem 91904, Israel}

\date{\today}

\begin{abstract}
The paper reassesses  the old but still controversial problem of the transverse force on a vortex and  the vortex mass. The transverse force from free bulk quasiparticles on the vortex, both  in the Bose and the Fermi liquid, originates from the Aharonov--Bohm effect. However, in the Fermi liquid one should take into account peculiarities of the Aharonov--Bohm effect for BCS quasiparticles described by two-component spinor wave functions. There is no connection between   the transverse force (either from free bulk quasiparticles or from  vortex-core bound quasiparticles) and the spectral flow in the vortex core in superfluid Fermi liquid, in contrast to widely known claims. 
In fact, there is no steady spectral flow in the core of the moving vortex, and the analogy with the Andreev bound states in the SNS junction, where the spectral flow is really possible, is not valid in this respect. 

 The role of the backflow on the vortex mass is clarified. The backflow is an inevitable consequence of a mismatch between the currents inside and outside the vortex core and restores the conservation of the particle number (charge) violated by this mismatch. In the Fermi liquid the backflow compensates the current through the core bound states, which is a source of the  vortex mass (the Kopnin mass). This results in renormalization of the Kopnin vortex mass by a numerical factor. 
   \end{abstract}

\maketitle


\section{Introduction}

Discussions and debates on the transverse force on a vortex in superfluids (neutral and charged)  continue during many decades and have been a topic of reviews and books \cite{RMP,Kop,KopR,Magn,ShelB,VolB}. They focused on quasiparticle contributions to this force, which are connected with geometrical (Aharonov--Bohm--Berry) phases in the superfluid around the vortex. Although in  most of practical cases the vortex can be considered as a massless object governed by the gyroscopic dynamics, the concept and the magnitude of the vortex mass was also vividly discussed in the Bose and the Fermi superfluids \cite{Suhl,VNP,KopM,Bay83,DL,Duan93,Duan,Son98,KopV,Kop,VolB}.

Despite a huge literature on this subject, there  still  remain some issues, which require further clarification, especially for the Fermi superfluids. The present paper addresses these issues. In particular, the paper rederives and discusses the origin of the transverse force from core states (the Kopnin--Kravtsov force) and the part of the transverse force from the free quasiparticles in the bulk of the Fermi superfluid, which seemed not to follow from simple semiclassical approach based on the Aharonov--Bohm effect. 
 \citet{Vol93} suggested that these transverse forces  originate from the spectral flow in the vortex core (see also Ch. 25 in his book \cite{VolB}). This interpretation was widely accepted \cite{Kop,KopR} and was a basis for the claim that the spectral flow in the vortex core presumably revealed in mutual friction measurements  experimentally models the cosmological baryogenesis in the early Universe \cite{BevN}. 
 
 The spectral flow concept is known both  in mathematics \cite{spFl}   and physics. According to its mathematical definition, the spectral flow is a number of eigenstates of an operator with eigenvalues  passing zero value at tuning of some parameter, on which the operator (and correspondingly its eigenstates) depends. A physical example of the spectral flow is the flow of the Andreev bound states  in  the ballistic Superconductor -- Normal metal -- Superconductor (SNS) junction\cite{stone96}. The energy of the Andreev state linearly depends on the superfluid phase difference between the superconductors forming the junction.   When the phase difference monotonously varies in time (the a.c. Josephson effect), the discrete energy level  cross the whole  superconducting gap passing the zero value of the energy. So in this example the parameter governing the spectral flow is the phase difference and the operator corresponds to the Bogolyubov-de Gennes  equations, which determine the Andreev bound states inside the gap. As was discovered long ago \cite{corSt,deGen}, the Andreev bound states exist also in cores of  vortices in Fermi superfluids, and Volovik argued that the process of vortex motion is accompanied by a steady shift of  core bound-state levels from negative-energy continuum to the positive-energy continuum, i.e., by the spectral flow across the superconducting gap similar to that in the SNS junction.  Any crossing of the gap by a bound state leads to transfer of the momentum, which leads to the transverse Kopnin--Kravtsov force. So momentum transfer from the vortex moving  with the relative velocity $\bm v_L-\bm v_n$ with respect to the normal component (or to impurities in superconductors) is realized not simply via jumps of particles between energy levels caused by collisions but via motion of energy  levels themselves in the energy space. 
  
This paper argues that the spectral flow cannot be responsible for any part of the transverse force simply because it is absent in a core of a moving vortex, as already was noticed by \citet{stone96} in the past. On the other hand, all kinds of transverse forces can be understood within common approaches like the scattering theory and the partial-wave expansion without any reference to the spectral flow. In particular, the part of the transverse force from  scattering of free bulk quasiparticles in the Fermi superfluid, which was presumed to originate from the spectral flow, directly follows from  peculiarities of the Aharonov-Bohm effect for BCS quasiparticles described by two-component spinor wave functions. Such  conclusions led to a necessity 
 to reassess Volovik's arguments in favor of the spectral flow  and to analyze why the analogy with the SNS junction, where the spectral flow definitely exists, is not applicable in this respect.
 
Addressing the vortex mass, the present paper revises different contributions to it,  compares them, and discusses  possible effects of the vortex mass on vortex motion. In particular, the paper analyzes  the so-called backflow vortex mass. The backflow mass is related with the kinetic energy of a superflow around the vortex core, which inevitably appears any time when the current density inside the core differs from that outside the core, and the intensity of the backflow is determined from the continuity of the total  fluid current.

Let us present the nomenclature of various forces, which enter the equation of vortex motion:
\be
mn_s \kappa[\hat z \times (\bm v_L-\bm v_s)]=\bm F_n +\bm F_c +{d \bm P \over dt}.
     \ee{vd}
The left-hand side is the Magnus force, which transfers momentum between the superfluid and the vortex. Here $\bm v_L$ and $\bm v_s$ are the vortex and the superfluid velocities,  $m$ is the particle mass, $n_s$ is the superfluid density, and $\kappa=h/m$ is the circulation quantum. The force
 \be
\bm F_n= -D (\bm v_L-\bm v_n)-D'[\hat z \times (\bm v_L-\bm v_n)]
     \ee{be}
transfers momentum between the normal component (the gas of free quasiparticles) and the vortex. Here $\bm v_n$ is the normal velocity.
The coefficients $D$ and $D'$,
\bem 
D=  {1 \over 3h^3} \int\frac{\partial
f_0(\varepsilon)}{\partial \epsilon} p^2 \sigma_\perp v_G \, d_3 \bm p,
\nonumber \\
D'=  {1 \over 3h^3} \int\frac{\partial
f_0(\varepsilon)}{\partial \epsilon} p^2 \sigma_\perp v_G \, d_3 \bm p
     \eem{D}      
are determined by the longitudinal (transport) and the transverse cross-sections $ \sigma_\parallel$ and $ \sigma_\perp$, which will be determined further in the paper.  Here $f_0(\epsilon)$ is the equilibrium Fermi distribution function of energy $\epsilon$ of free quasiparticles, and $v_G$ is the projection of the quasiparticle group velocity on the plane normal to the vortex line. The transverse force proportional to $D'$ is the Iordanskii force.

The force $\bm F_c$ transfers momentum from the quasiparticles occupying bound states in the vortex core to impurities in superconductors or to free bulk quasiparticles constituting the bulk normal component of the $^3$He superfluid. The force has also two components, longitudinal and transverse to the relative normal velocity $\bm v_n-\bm v_L$, the latter called the Kopnin--Kravtsov force \cite{Kop76b}.

Finally, $d\bm P/dt$ is the inertial force, which is a product of the vortex mass and the vortex acceleration $d\bm v_L/dt$, the momentum $\bm P$ being the momentum of the vortex dependent on $\bm v_L$. 

The  theory presented in this paper assumes that the quasiparticle mean-free path is much longer that the core size and therefore it cannot be used for high temperatures. The whole paper addresses neutral superfluids, although the results for the Fermi superfluids are relevant also for type II $s$-wave superconductors, since  the effects of magnetic fields usually are not essential for vortex dynamics \cite{Kop}. In superconductors the normal velocity $\bm v_n$ usually vanishes in the coordinate frame related to the crystal lattice. 

The paper starts from Sec. \ref{GO} reminding the old results for semiclassical scattering of quasiparticles by a vortex. This shows the connection of the transverse force with the Aharonov--Bohm effect for quasiparticles.  
Section~\ref{BCS} considers the scattering of BCS quasiparticles on the basis of the Bogolyubov-de Gennes  equations. The analysis is done using the geometric optics and the partial-wave method. It demonstrates that the whole transverse force from free bulk quasiparticles is fully explained by the Aharonov--Bohm effect  without referring to the concept of spectral flow. But one must take into account the peculiarities of the  Aharonov--Bohm effect for BCS quasiparticles described by two-component spinor wave functions.  Section~\ref{corSt} reminds  properties of bound states in the  vortex core in the Fermi superfluid focusing on the role of superfluid motion outside the core. Sections~\ref{vmB} and \ref{fermiM} consider various contributions to the vortex mass in the Bose and the Fermi liquid respectively. Section~\ref{KE} discusses the derivation of the transverse force and the vortex mass from the Boltzmann equation focusing on the effect of superfluid transport past the vortex and on the comparison with the analysis of the previous sections. Section~\ref{VorIn} analyzes possible effects of the vortex mass on vortex dynamics.  Concluding discussion of the results and the shortcomings  of the spectral flow interpretation of the transverse force is presented in Sec.~\ref{spFl}. Two appendices address more special issues:  the simplified derivation of the spectrum of bound states for a core with linear growth of the gap as a function of the distance from the axis (App.~\ref{bss}) and the derivation of the vortex mass for a core with linear growth of density in the Bose superfluid (App.~\ref{bm}).

\section{Transverse force from the semiclassical scattering theory (geometric optics)} \label{GO}

It is useful to start from the simplest approach to this problem based on the  semiclassical scattering theory, which   was first used for rotons  by \citet{Lif57} long ago.

The theory is based on the geometric optics. A quasiparticle moves along a well-defined trajectory and its motion is described by variation of the position vector $\bm  R$ and the  momentum $\bm  p$ of the quasiparticle in time.
The classical Hamilton
equations for them are:
\be
\frac{d\bm  R}{dt}= \frac{\partial \epsilon}{\partial \bm  p},~~~
\frac{d\bm  p}{dt}=- \frac{\partial \epsilon}{\partial \bm  R}.
                                      \ee{eq:r1}
Here
\be
\epsilon(\bm  p) = \epsilon_0(\bm  p) + \bm  p\cdot \bm  v_v
                                           \ee{eq:r2}
is the energy of the quasiparticle in the moving fluid,
$\epsilon_0$ is the quasiparticle energy in the resting fluid,  and $\bm  v_v$ is the velocity induced by a
rectilinear vortex:
\be
\bm  v_v = \frac{[\kappa \times \bm  r]}{2\pi r^2},
                              \ee{eq:r3}
where $\bm  r$ is a position vector in the plane normal to the
vortex line (the projection of $\bm  R$ on that plane). In order to simplify discussion we assume that the quasiparticle moves in the normal plane, so its momentum $\bm p$ lies in this plane.

\begin{figure}
\includegraphics[width=.5\textwidth]{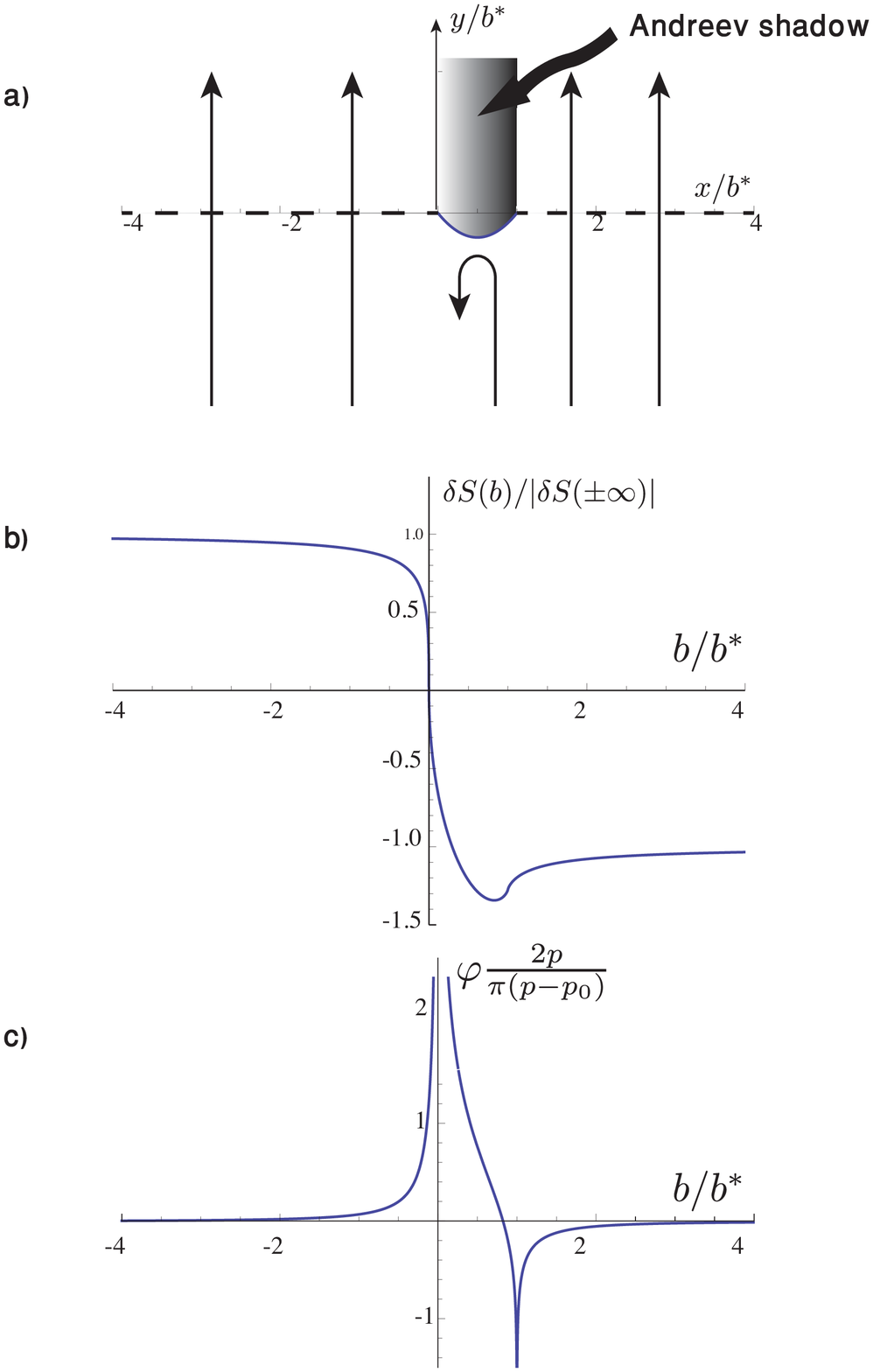}
\caption[]{(Color online) Semiclassical scattering of rotons. a) Rotons at trajectories with impact parameters   $b=x<0$ or $b>b^*$ move past the vortex.  Rotons at trajectories with  $0<b<b^*$ are fully reflected by the vortex. 
The shaded area  (Andreev shadow) is classically forbidden for rotons. b) Action variation $\delta S(b)$ along the trajectory as a function of the impact parameter $b$ (dimensionless variables). c) Scattering angle $\varphi(b)$ as a function of the impact parameter $b$ (dimensionless variables).}
\label{f11-00}
\end{figure}

The vortex velocity field produces a force $\bm  \nabla(\bm 
p \cdot \bm  v_v)$ on the quasiparticle. The  force may be considered as weak and the quasiparticle
trajectory  as nearly rectilinear. Suppose that the trajectory is
parallel to the $y$-axis (Fig.~\ref{f11-00}a) and its impact parameter (the distance between
the vortex line and the trajectory) is $b=x$. Then \eq{eq:r1} gives
\be
\frac{dy}{dt}=v_G,~~~\frac{d\bm  p}{dt}=-\bm  \nabla (\bm  p\cdot \bm 
v_v) .                                 \ee{eq:r4}
Here  $\bm v_G=\partial \epsilon_0(\bm  p)/\partial \bm  p$ is the
quasiparticle group velocity in the resting fluid, which is in our case approximately  parallel to the axis $y$. Excluding time
from these equations one has a differential equation determining
the quasiparticle momentum variation along the trajectory:
\be
\frac{d\bm  p}{dy}= -\frac{1}{v_G}\bm  \nabla (\bm  p\cdot \bm  v_v).
                             \ee{eq:r5}
Integration of this equation assuming that the group
velocity $v_G$ does not vary along the trajectory yields
\be
\bm  p(y) = \bm  p - \frac{p}{v_G}\bm  v_v(b,y),
                                   \ee{eq:r6}
where $\bm p=\bm  p(-\infty)$ is the momentum at $y=-\infty$. 

The scattering angle $\varphi$ between the final and the initial momenta of the
quasiparticle determines the momenta $p(1-\cos \varphi)$  and $p\sin\varphi$, which are longitudinal and transverse with the respect to the incident momentum $\bm p$. The momenta are  transferred by the scattered quasiparticle to the vortex. Correspondingly, the longitudinal and the transverse forces on the vortex from quasiparticles [see Eqs.~(\ref{be}) and  (\ref{D})] are determined by the longitudinal (transport),  
\be
\sigma_\parallel=\int \limits_{-\pi}^\pi \sigma(\varphi)(1 - \cos \varphi)d\varphi \approx
\int \limits_{-\infty}^\infty \frac{\varphi(b)^2}{2}db,
                                  \ee{longG}
and the transverse, 
\bem
\sigma_\perp=\int \limits_{-\pi}^\pi \sigma(\varphi)\sin\varphi\,d\varphi\approx 
\int \limits_{-\infty}^\infty \varphi (b)  db,
    \eem{tranG} 
effective cross-sections. Here 
\be
  \sigma(\varphi)=\frac{db}{d\varphi}
                                        \ee{eq:r13}
is the differential cross-section. In our analysis we assume that the scattering angle $\varphi \approx -p_x/p$ is small.  

In the Hamilton--Jacobi theory the momentum is connected with the classical
action: $\bm p = \partial S /\partial \bm r$. Then $ p_x =
\partial \delta S (b)/ \partial b$,  where 
\begin{equation}
\delta S(b) =\int_{-\infty}^{\infty} [p(y) -p] dy=   - {p  \over v_G }\int_{-\infty}^{\infty} \nabla_y v_{vy}dy 
       \label{S} \end{equation}     
is the variation of the classical
action along the trajectory, which  is a function of the impact parameter $b$. This yields:
\bem
\sigma_\perp=-{1\over p}\int \limits_{-\infty}^\infty  {\partial \delta S(b)\over \partial b} db
= \frac{ \delta S(-\infty) - \delta S(+\infty)}{p}.
    \eem{act-sigma} 
Bearing in mind
that the velocity induced by the vortex is $\bm  v_v= (\kappa/2\pi) \bm 
\nabla \phi(\bm  r)$ where the phase $\phi = \arctan{(y/x)}$ is the
azimuthal  angle for the two--dimensional position vector $\bm  r$
[see \eq{eq:r3}] one obtains that 
\begin{equation}
\delta S(b) =   - {p\kappa   \over 2\pi v_G }\int_{-\infty}^{\infty} \frac{b}{b^2+y^2}dy = -\mbox{sign} b {p\kappa   \over 2  v_G }.
       \label{Sa} \end{equation}     
Eventually \eq{act-sigma} yields the transverse  cross-section\cite{Son75}
\bem
\sigma_\perp = {\kappa \over v_G}.
    \eem{cl-sigma} 
So we have obtained for the transverse cross-section an universal expression, which looks valid for any quasiparticle spectrum. The cross-section is proportional to the total variation of the classical action around the vortex line. Because of correspondence of the classical action  to the quantum mechanical  phase, this points out connection of the transverse force with the geometric phase, or the Aharonov--Bohm effect \cite{Son75}. \Eq{cl-sigma} yields a correct transverse cross-section for phonons  even though the
semiclassical theory is not valid for phonons: there is no well-defined
classical trajectories for phonons except for large impact parameters  $b$ at
which the scattering angle $\varphi$ is negligible. In the case of phonons the group velocity $v_G$ is the sound velocity $c_s$.

The simple expression (\ref{act-sigma}) for the transverse cross-section  does not depend on how the scattering angle varies as a function of the impact parameter  because an integrand in \eq{act-sigma} is a derivative of the action. But it does assume that the action is a continuous  function of the impact parameter. Now we shall check it for rotons in superfluid $^4$He. 

 The energy spectrum for rotons is $\epsilon_0(\bm 
p)=\Delta +(p-p_0)^2/2\mu$, where $\Delta$ is the roton gap and $\mu$ is the roton mass. According to the energy conservation
law following from the Hamilton equations, \eq{eq:r1},
one has:
\be
\Delta +\frac{[p(y)-p_0]^2}{2\mu} + \bm  p(y) \bm  v_v(y) = \Delta +\frac{(p-p_0)^2}{2\mu}.
                             \ee{eq:r7}
The right-hand side is the energy far from the vortex line, where $p=p(-\infty)$. The variation of the roton group velocity along the trajectory
with the impact parameter $b$ is given by
\bem
v_G(y)=\frac{p(y)-p_0}{\mu}=\frac{1}{\mu}\sqrt{(p-p_0)^2-
2\mu v_{vy}}
\nonumber \\
= v_G\sqrt{1- \frac{bb^*}{b^2+y^2}}.
                                 \eem{eq:r8}
Here the characteristic scattering length
\be
b^*=\frac{\kappa \mu  p}{\pi (p-p_0)^2}
                                   \ee{eq:r9}
is introduced and the $v_G=v_G(-\infty)=(p-p_0)/\mu$ is the
roton group velocity far from the vortex line. The
asymptotic expression \eq{eq:r6} obtained at constant $v_G$ is
valid for large impact parameters $|b| \gg b^*$. 

In the classical scattering theory the point $y=0$ on the trajectory is a turning point:  At $y<0$ the quasiparticle approaches to the scattering center (vortex line in our case) while at $y>0$ the quasiparticles moves away from the vortex line. Equation (\ref{eq:r8}) shows that for   impact parameters $0<b<b^*$ the quasiparticle cannot reach the usual turning point $y=0$ since at $y =-y^*$,  where $y^*= \sqrt{b^*b-b^2}$, the group velocity vanishes, and the quasiparticle starts to move back to $y=-\infty$ without an essential change of its momentum. This is Andreev reflection well known in the theory of superconductivity. At the point $y =-y^*$  $p=p_0$ and  the transition between two branches
of the roton spectrum with $p>p_0$  (positive branch, parallel momentum and group velocity) and $p<p_0$  (negative branch, antiparallel momentum and group velocity) occurs.  Due to the
Andreev reflection the shadow region is formed near the vortex line
which is not available for the roton classical trajectories. That
shadow (Andreev shadow)  region is shown in Fig.~\ref{f11-00}a.

Let us find the classic action variation  first for trajectories with impact parameters $b>b^*$ or $b<0$, when   there is no Andreev reflection and the incident roton  stays at the same branch after the collision.
Taking into account variation of the roton group velocity along the trajectory, \eq{eq:r8}, the variation of the action along the trajectory for the incident momentum $p>p_0$  is
\bem
\delta S(b)=\int\limits_{-\infty}^\infty  (p(y)-p) \,dy
\nonumber \\
=(p-p_0)\int\limits_{-\infty}^\infty \left(\sqrt{1 -{b^*b\over b^2+y^2}}-1\right)dy
\nonumber \\
=2 \mbox{sign}(b) (p-p_0)\left[\left(b-b^*\right) F\left(b^*\over b\right)- b E\left(b^*\over b\right) \right],~~~
   \eem{S+}
where 
\bem 
F(m)= \int_0^{\pi/2} {d\theta\over  \sqrt{1 -m\sin^2\theta }},
\nonumber \\
E(m)= \int_0^{\pi/2} \sqrt{1 -m\sin^2\theta}d\theta
  \eem{FE}
are complete elliptic integrals of the first and the second order respectively. In the limits $b\to \pm \infty$ \eq{S+} reduces to \eq{Sa}. 

In the interval $0<b<b^*$ trajectory ends at the Andreev reflection point with the coordinate $y=-y^*$. The incident roton with momentum $p=p_0+(p-p_0)>p_0$ returns after the  Andreev reflection to $y=-\infty$ at the other branch with the same energy but a slightly different momentum $p_- =p_0 -(p-p_0)<p_0$.
The variation of the action along the whole path is 
\bem
\delta S(b)=\int\limits_{-\infty}^{-y^*}  p(y) \,dy + \int\limits^{-\infty}_{-y^*}  p_-(y) \,dy       -\int\limits_{-\infty}^a  p\,dy-\int\limits^{-\infty}_{a}  p_- \,dy
\nonumber \\
=2(p-p_0)\left[\int\limits_{-\infty}^{-y^*}  \left(\sqrt{1 -{b^*b\over b^2+y^2}}-1\right)dy - y^*-a\right].
\nonumber \\
   \eem{}
Here $a$ is an undefined constant, which does not depend on $b$ and therefore has no effect on the scattering angle $\varphi$. Choosing $a=0$ one eliminates any discontinuity of $S(b)$ at $b=0$ and $b=b^*$. \footnote{If one chooses nonzero $a$, discontinuity of $S(b)$ at $b=0$ and $b=b^*$ leads to two $\delta$-function contributions of opposite signs in the angle $\varphi$, which cancel each other in the transverse cross-section $\sigma_\perp$.}
Introducing the angle variable again one obtains the expression  
\bem
\delta S(b)
=2  (p-p_0)\left[\left(b-b^*\right) F\left(\phi, {b^*\over b}\right)- b E\left(\phi, {b^*\over b}\right) \right]
\nonumber \\   \eem{}
in terms of incomplete elliptic integrals
\bem 
F(\phi,m)= \int_0^{\phi} {d\theta\over  \sqrt{1 -m\sin^2\theta }},
\nonumber \\
E(\phi,m)= \int_0^{\phi} \sqrt{1 -m\sin^2\theta}d\theta,
  \eem{FEin}
where $\phi=\arcsin \sqrt{b/b^*}$.

In Figs.~\ref{f11-00}b and \ref{f11-00}c the action $\delta S(b)$ and the scattering angle $\varphi(b)=-\partial \delta S(b)/\partial b$ are plotted as functions of the impact parameter $b$ (in dimensionless variables). The angle $\varphi$ has weak singularities at $ b=0$ and $b=b^*$, which are integrable in the integral  for the transverse cross-section $\sigma_\perp$ [\eq{tranG}].

Substituting the transverse cross-section (\ref{cl-sigma}) into the expression for the parameter $D'$, which determines the Iordanskii force [see \eq{D}], one obtains $D'=-\kappa m n_n$, where $n_n$ is the normal particle density.
This rather simple and universal expression  tempts to claim its universal topological origin,
since $\kappa$ in this expression is a topological charge. However, in the
Sec.~\ref{BCS} we shall see that the expression is not universal. For
quasiparticles in a BCS superconductor with energy much exceeding the gap an
additional small factor appears in this expression.

\section{Scattering of bulk free  BCS quasiparticles by a vortex  in Fermi superfluids} \label{BCS}

\subsection{The Bogolyubov-de Gennes  equations}

The wave function of quasiparticles in the BCS theory is a spinor with two components,
\begin{equation}
\psi(\bm  R) = \left( \begin{array}{c} u(\bm  R)\\  v(\bm  R) \end{array}
\right),
     \label{spinor} \end{equation}
which are determined from the Bogolyubov-de Gennes
equations \citep{deGen}:
\begin{equation}
-{\hbar^2 \over 2m} \left(\bm  \nabla^2 + k_F^2\right) u (\bm  R) + 
\Delta(\bm R) e^{i\theta(\bm  R)}  v(\bm  R) =\epsilon u(\bm  R),
     \label{BGu} \end{equation}
\begin{equation}
{\hbar^2 \over 2m}  \left(\bm  \nabla ^2 + k_F^2\right) v (\bm  R) + \Delta(\bm R) 
e^{-i\theta(\bm  R)}  u(\bm  R) =\epsilon v(\bm  R).
     \label{BGv} \end{equation}
Here $k_F$ is the Fermi wave number, and the gap $\Delta(\bm R)$ can vary in  space. 
The equations correspond to the Hamiltonian with the density 
\bem
{\cal H}={\hbar^2 \over 2m} ( |\nabla u|^2 - k_F^2  |u|^2)-{\hbar^2 \over 2m} ( |\nabla v|^2 - k_F^2  |v|^2)
\nonumber \\
+\Delta(\bm R) e^{i\theta(\bm  R)}u^* v+\Delta(r) e^{-i\theta(\bm  R)}v^* u.
   \eem{ham}

If a superfluid is at rest the order parameter phase $\theta$ is a
constant and the solution of  the Bogolyubov-de Gennes
equations is a plane wave
\be 
\left(  \begin{array}{c}
u_0    \\
  v_0 e^{i\theta}     
\end{array}  \right)e^{i \bm k \cdot \bm R},
   \ee{pwsol}
where
\be
\left(  \begin{array}{c}
u_0    \\
  v_0      
\end{array}  \right)=\left(  \begin{array}{c}
\sqrt{{1\over 2} \left( 1+  {\xi\over \epsilon_0}\right)}    \\
  \sqrt{{1\over 2} \left( 1-  {\xi\over \epsilon_0}\right)}       
\end{array}  \right).
          \ee{u0v0}
The energy is given by the well-known BCS quasiparticle spectrum
$\epsilon_0= \pm \sqrt{\xi^2 + \Delta^2}$. Here  $\xi = ({\hbar^2 / 2m})(k^2 - k_F^2)\approx \hbar v_F (k-k_F)$  is
the quasiparticle energy in the normal Fermi-liquid and $v_F= k_F/m$ is the Fermi velocity. The two  wave numbers 
\be
k_\pm =\sqrt {k_F^2\pm 2 \sqrt{\epsilon_0^2-\Delta^2}}
   \ee{kpm}
   correspond to the particle-like (+) and the hole-like (-) branches of the spectrum.

The  Bogolyubov-de Gennes equations are written for the wave function of quasiparticles, and, as in the case of the Schr\"odinger  equation, there is the continuity equation for the probability $|u|^2+|v|^2$ to find the quasiparticle in some point in space:
\be
{\partial (|u|^2+|v|^2)\over dt}=- \bm \nabla \cdot \bm g, 
   \ee{pqu} 
where
\be
\bm g = -{i\hbar \over 2m } (u^*\bm\nabla u -u\bm\nabla u^*)+{i\hbar \over 2m }(v^*\bm\nabla v -v\bm\nabla v^*)
  \ee{}
is the probability flux. \Eq{pqu} is a manifestation of the conservation law for the number of quasiparticles. But the number of quasiparticles and the probability flux $\bm g $ are not the same as the number of particles (charge) and the particle current $\bm j$. The Hamiltonian of \eq{ham} is not gauge-invariant and there is no conservation law for the particle number. The Bogolyubov-de Gennes equations lead to the following equation for time variation of the particle density $|u|^2-|v|^2$:
\be
{\partial (|u|^2-|v|^2)\over dt}=- \bm \nabla \cdot \bm j +2i\Delta (e^{-i\theta}v^*u-e^{i\theta}vu^*), 
   \ee{pp} 
where
\be
\bm j = -{i\hbar \over 2m} (u^*\bm\nabla u -u\bm\nabla u^*)-{i\hbar \over 2 m}(v^*\bm\nabla v -v\bm\nabla v^*)
  \ee{cur}
is the particle current.  \Eq{pp} contains a source (the last term in the right-hand side) related with possible changing of the total particle number. Globally the number of particles is  of course a conserved quantity. The source in the continuity equation for the particle density corresponds to conversion of the superfluid part of the liquid to the normal  one and vice versa in inhomogeneous states. In order to restore the global conservation law one should solve the  Bogolyubov-de Gennes equations together with the self-consistency equation for the order parameter proportional to the gap. This property of the  Bogolyubov-de Gennes equations is well known in the theory of superconductivity  \citep{BTK}.

 \subsection{Superfluid motion in the  Bogolyubov-de Gennes equations}
 
 Superfluid velocity is determined by the order parameter phase gradient:
 \be
 \bm v_s={\kappa_c\over 2\pi} \bm \nabla \theta,
      \ee{}
where $\kappa_c=h/2m$ is the circulation quantum  for the Cooper-pair condensate and $m$ is the particle mass.
 Assuming constant absolute value of the gap $\Delta$ and gradient of phase, the solution of the  Bogolyubov-de Gennes equations is
 \be
\left(  \begin{array}{c}
u    \\
  v      
\end{array}  \right)=\left(  \begin{array}{c}
u_0  e^{i(\bm k+ \bm\nabla \theta_1)\cdot \bm R}    \\
 v_0 e^{i(\bm k-\bm\nabla \theta_2)\cdot \bm R}       
\end{array}  \right)
       \ee{vs-sol}
Here we introduced separate phases $\theta_1$ and $\theta_2$  for two spinor components. Their sum determines the order parameter phase $\theta=\theta_1+\theta_2$.   The spinor (\ref{vs-sol}) corresponds to the energy (neglecting terms of the second order in phase gradients)
\bem
\epsilon= \epsilon_0(k)+{\hbar \kappa_c \over 2\pi}\bm  k \cdot \bm \nabla \theta+ {\partial \epsilon_0\over \partial \bm  k} \cdot {\bm \nabla \theta_1-\bm\nabla \theta_2\over 2}
\nonumber \\
=\epsilon_0(k) +\hbar \bm  k \cdot 
\left[\bm v_s+{ \xi \over  \epsilon_0 }{\kappa_c\over 2\pi} (\bm \nabla \theta_1-\bm\nabla \theta_2)\right].  
        \eem{vs-spec}
It looks as if the  phase difference $\theta_1-\theta_2$ were  of no importance since it can be removed by redefinition of the wave vector $\bm k$. Choosing $\theta_1= \theta_2$ one obtains the expression for the quasiparticle energy  following from the Galilean invariance and well known from textbooks on superconductivity \cite{deGen}: $\epsilon= \epsilon_0 +\hbar \bm  k \cdot \bm v_s$. But another choice is required in the theory of quasiparticle scattering by a vortex:  either $\theta_1=0$ or $\theta_2=0$. This is dictated by cyclic boundary conditions for spinor components on the closed path around the vortex (see Secs.~\ref{SA} and \ref{PW}).

For the choice $\theta_1= \theta_2$  \eq{cur} yields the following expression for the current in the plane-wave state:
\be
\bm j ={\hbar \bm k \over m } +N(\bm k)\bm v_s.
  \ee{cur1}
So the superfluid velocity contribution to the current is proportional to the charge $N(\bm k)=|u_0|^2-|v_0|^2$  in the state.

 \subsection{Scattering of free BCS quasiparticles by a vortex: simple approach} \label{SA}
 
  The mutual friction force has  been calculated for
pure type II superconductors long ago \citep{Kop76,Gal}. Since the BSC theory describes also the superfluid $^3$He and the effect of the magnetic field is insignificant for mutual friction in type II superconductors these calculations are relevant also for singular vortices in  the superfluid $^3$He. In this subsection we use  simple approaches: geometric optics for low energies $\epsilon_0-\Delta  \ll\Delta$ and perturbation theory for high energies $\epsilon_0 \gg \Delta$. A more accurate theory based on the partial-wave expansion will be considered in the next subsection.

When the energy of the quasiparticles is close to the energy gap of the
superconductor  ($\xi \ll \Delta$), the BCS quasiparticle spectrum $\epsilon_0\approx \Delta + v_F^2 \hbar^2 (k-k_F)^2 /2\Delta$  is identical to the roton spectrum with the roton minimum momentum replaced by the Fermi momentum $\hbar k_F$ and the roton mass $\mu$ replaced by $\Delta/v_F^2$, where $v_F=\hbar k_F/m$ is the Fermi velocity. So  the semiclassical theory for rotons  can be directly applied to such BCS quasiparticles, and the transverse  cross-section for them is given by Eq. (\ref{cl-sigma}), in which the circulation quantum $\kappa$ is replaced by  the circulation quantum $\kappa_c=h/2m$ for the Cooper-pair condensate and the group velocity for the BCS quasiparticles is $v_G=v_F \xi/\epsilon_0 $.  Figure~\ref{f11-00} illustrating scattering of rotons by the vortex is relevant also for low-energy quasiparticles scattered by the vortex. The phenomenon of the nearly 180\% reflection of quasiparticles from the area of the Andreev shadow shown in the figure is important for description of zero-temperature superfluid turbulence \cite{FP,BarAnd}.

If the quasiparticle energy is much larger than the
superconducting gap, the group velocity $v_G$ approaches to the Fermi velocity $v_F$ and  the method of classical trajectories yields the transverse cross-section  $\kappa_c/v_F$. This result does not look reasonable, because the cross-section  being small still does not vanish  in the limit $\Delta \to 0$. Indeed, the partial-wave calculations  \citep{Kop76,Gal} yielded that  in the limit of small $\Delta/\xi$ the transverse cross-section
differed from the semiclassical result of Eq.  (\ref{cl-sigma}) by the
factor $\Delta^2/2\xi^2$. This also followed from the solution of the Bogolyubov-de Gennes
equations in the Born approximation \citep{PRB7} as shown below.

Let us consider the perturbation theory with respect to the gap $\Delta$ and the superfluid velocity $\bm v_s= (\kappa_c /2\pi)\bm \nabla \theta$. For the sake of simplicity the wave vector $\bm k$ lies in the plane normal to the vortex axis. In our case the superfluid velocity is the velocity $\bm v_v$ induced by the vortex. In the
zero-order approximation $u\sim  \exp(i\bm  k \cdot \bm  r)$ and $v=0$. In
the first-order approximation the second  Bogolyubov-de Gennes equation
(\ref{BGv}) yields 
\begin{eqnarray} 
v =\left\{\frac{\Delta
\exp(-i\theta)}{\xi(k) + E(k)} +\frac{\Delta \exp(-i\theta)}{[\xi(k) +
E(k)]^2}{\hbar^2 \over m}(\bm  k \cdot \bm  \nabla \theta)\right\} 
e^{i\bm  k \cdot \bm  r}. \nonumber \\ 
      \label{pert-v} \end{eqnarray} 
The first term in curled brackets yields a correction to the quasiparticle
energy $\propto \Delta^2$, but does not contributes to scattering which is
determined by the order-parameter phase gradients. So we keep only the second
term proportional to $\bm  \nabla \theta$. Inserting it to the first
Bogolyubov-de Gennes equation (\ref{BGu}) one obtains the following equation
for the  correction $u'$ to the quasiparticle amplitude $u\sim 1$:
\begin{equation}
(\nabla^2 + k^2) u' =  (\bm  k \cdot \bm  \nabla \theta)
\frac{\Delta^2}{2\xi^2} e^{i\bm  k \cdot \bm  r}.
       \label{pert-u} \end{equation}
This equation is similar to the wave equation  for the sound wave propagating pass the vortex \cite{PRB7,Magn} and using
this analogy one easily obtains the expression for the
transverse cross-section:
\begin{equation}
\sigma_\perp = \frac{\Delta^2}{2\xi^2} \frac{\pi}{k_F}
 = \frac{\Delta^2}{2\xi^2} \frac{\kappa_c}{v_F}.
       \label{sig-S-el} \end{equation}
The cross-section vanishes at $\Delta \to 0$ as expected. But the question where the geometric optics went wrong still remains.   The answer is that the cyclic boundary conditions were violated with the choice $\theta_1=\theta_2 =\theta/2$. Let us move the spinor given by \eq{vs-sol} along a closed path around the vortex line. After closing the path the phase $\theta$ obtains the shift $2\pi$ but the shifts of the phases $\theta_1 $ and $\theta_2$ are equal to $\pi$. So the periodic boundary conditions for the spinor components $u$ and $v$ are  violated. They are satisfied only if either $\theta_1$ or $\theta_2$ vanishes. According to \eq{vs-spec} this modifies the expression for the quasiparticle energy in the vortex velocity field:
\bem
\epsilon=\epsilon_0(k) +(\hbar \bm  k\pm m \bm v_G) \cdot \bm v_v.  
        \eem{}
Then the value of $p$ in \eq{Sa} must be replaced by $\hbar k\pm m  v_G$. Choosing - for quasiparticles and + for quasiholes (this is dictated by a physically reasonable condition that the cross-section vanishes far from the Fermi surface) one obtains the transverse cross-section \cite{Kop76,Gal}
\be
\sigma_\perp={\kappa_c \over v_G}-{\kappa_c \over v_F}={\kappa_c \over v_F}\left({ \epsilon_0 \over \sqrt{\epsilon_0^2-\Delta^2}} -1\right).
         \ee{gopt}
In the limit $\epsilon_0 \gg \Delta$ this agrees with the expression (\ref{sig-S-el}) obtained from the perturbation theory with respect to $\Delta$. A more rigorous partial-wave analysis of the next subsection  confirms this result for any ratio $ \Delta/\epsilon_0$. This provides an explanation for shortcoming of the naive geometric-optics analysis: It ignored peculiarities of the Aharonov--Bohm effect for BCS quasiparticles and used an improper
definition for the quasiparticle phase shift along the trajectory.   We shall continue the discussion of this issue in the end of the next subsection.

\subsection{Partial-wave analysis of scattering of free BCS quasiparticles by a vortex} \label{PW}

The partial-wave analysis in the cylindric coordinates $r,~\phi, ~z$ uses expansion of the spinor components in eigenfunctions $e^{il\phi}$ of the orbital moment. In the presence of a vortex the phase of the order parameter $\Delta e^{i\theta}$ around the vortex   is $\theta=\phi$ and the partial wave expansion for the  wave function is 
\be
u=\sum_l u_l e^{il\phi},~~v=\sum_l v_l e^{i(l-1)\phi},
    \ee{}
where $u_l$ and $v_l$ must satisfy the Bogolyubov--de Gennes equations for  partial waves:
\bem
-{\hbar ^2\over 2 m }\left({d^2u_l\over dr^2}+{1\over r}{du_l\over dr}-{l^2u_l\over r^2}\right) +\Delta u_l
=\left(\epsilon + {\hbar ^2k_F^2\over 2m}\right)v_l,
\nonumber \\
{\hbar ^2\over 2 m }\left({d^2v_l\over dr^2}+{1\over r}{dv_l\over dr}-{(l-1)^2v_l\over r^2}\right)+\Delta u_l
=\left(\epsilon - {\hbar ^2k_F^2\over 2m}\right)v_l. \nonumber \\
  \eem{BdGlv}
In our case the orbital number $l$ is not an ideal quantum number since the two components of the spinor correspond to two different orbital numbers $l$ and $l-1$. 

In order to find the scattering phases, we shall look for the  semiclassical solution of the Bogolyubov--de Gennes equations
for the scaled amplitudes $U_l=u_l \sqrt{r}$ and  $V_l=v_l \sqrt{r}$:
\bem
-{\hbar ^2\over 2 m }\left({d^2U_l\over dr^2}-{l^2-1/4\over r^2}U_l\right) +\Delta V_l=\left(\epsilon + {\hbar ^2k_F^2\over 2m}\right)U_l,
\nonumber \\
{\hbar ^2\over 2 m }\left({d^2V_l\over dr^2}-{(l-1)^2-1/4 \over r^2}V_l \right)+\Delta U_l=\left(\epsilon - {\hbar ^2k_F^2\over 2m}\right)V_l.
\nonumber \\
  \eem{BdGb}
The semiclassical solution of the Bogolyubov--de Gennes equations (\ref{BdGb}) for  partial waves is 
\be
\psi \sim {1\over \sqrt{k_\pm }}\left(  \begin{array}{c}
\sqrt{{1\over 2} \left( 1+  {\sqrt{\epsilon_0^2-\Delta^2}\over \epsilon_0}\right)}    \\
  \sqrt{{1\over 2} \left( 1- {\sqrt{\epsilon_0^2-\Delta^2}\over \epsilon_0}\right)}       
\end{array}  \right)e^{i\int^rk_\pm(r) dr} ,
          \ee{}
where $\epsilon_0=\epsilon -(l-1/2)/2r^2$ and
\bem
k_\pm(r)^2=k_F^2-{(l-1/2)^2 \over  r^2}\pm 2\sqrt{ \left(\epsilon-{l-1/2\over 2r^2}\right)^2-\Delta^2}.
\nonumber \\
  \eem{wkb}
If a quasiparticle with the wave number  $k_+$  is incident on the vortex line, it will be reflected either as a quasiparticle with the same number $k_+$ (usual reflection)  or  as a quasiparticle belonging to the hole-like branch with $k_-<k_F$ (Andreev reflection).  The usual reflection occurs at the turning point  determined by the condition $k_+(r_r)=0$. In the Andreev reflection point $r=r_a$ the inner radical in \eq{wkb} vanishes, i.e., $\epsilon-{l-1/2\over 2r_a^2}\pm \Delta=0$.  
The type of the reflection depends on which turning point is reached earlier: usual or Andreev reflections take  place if $r_r>r_a$ or   $r_r<r_a$ respectively.

 In the following we shall look for the wave function for large orbital numbers $l$, which correspond to large impact parameters. Then only usual reflection is possible, and one can expand the inner radical in \eq{wkb} with respect to $(l-1/2)/r^2$. Then 
\bem
k^2\approx k_\pm ^2-{(l-1/2)^2\pm (l-1/2)\epsilon/\sqrt{ \epsilon^2-\Delta^2} \over  r^2}
\nonumber \\
\approx k_\pm ^2-{(l-1/2\pm \epsilon/2\sqrt{ \epsilon^2-\Delta^2} )^2\over  r^2},
  \eem{wkb1}
where   $k_\pm$ are values of   $k_\pm(r)$ at $r\to \infty$. The total phase accumulated after quasiparticle motion from very large $r$ to the turning point and back to large $r$ is 
\bem
\Phi_l=2 \int_{r_t}^r \sqrt{k_\pm ^2-{(l-1/2\pm \epsilon/2\sqrt{ \epsilon^2-\Delta^2} )^2\over  r^2}}dr
\nonumber \\
 -{\pi\over 2}=  2 k_\pm r-\pi\left|l-{1\over 2}\pm {\epsilon\over 2\sqrt{ \epsilon^2-\Delta^2}}\right|-{\pi\over 2}.
  \eem{phiAc}
Here the phase shift $-\pi/2$ originates from the close  vicinity of the reflection point where the semiclassical approach becomes invalid.\cite{LLqu} In order to find the phase shift from scattering of the quasiparticle (particle branch, the upper sign in the expressions above) by the vortex  one should  subtract the phase shift  $\Phi_{l0} = 2k_+ r- \pi (|l|+1/2)$ of the $l$-partial  wave function in the uniform state. This follows from the expansion of a plane wave in partial waves.  Then  the scattering phase shift  is
\bem
\delta_l ={\Phi_l-\Phi_{l0}\over 2}=-{\pi\over 2}\left|l-{1\over 2}+ {\epsilon\over 2\sqrt{ \epsilon^2-\Delta^2}}\right| +|l|{\pi\over 2}
\nonumber \\
={\pi\over 4}\left(1- {\epsilon\over \sqrt{ \epsilon^2-\Delta^2}}\right)\mbox{sign}l.
      \eem{}
The variation of the classical action along the quasiparticle trajectory is connected with the quantum-mechanical scattering phase shift by the relation $\delta S(b) =2\hbar \delta _l$, where $b\approx l/k_F$. Thus one obtains that
\be
\delta S(\pm\infty)= \mp {\hbar k_F \kappa_c \over 2}\left({1\over v_G}-{1\over v_F}\right).
  \ee{SAB}
Inserting it into \eq{act-sigma}  yields the transverse cross-section  \eq{gopt}  obtained after correction of the geometric-optics expression.  In the case of the hole branch (the lower sign in the expressions above) one should subtract from the phase shift  $\Phi_l $ in the vortex state the phase shift $\Phi_{(l-1)0} = 2k_- r- \pi (|l-1|+1/2)$ of the $(l-1)$-partial  wave function in the uniform state.

The second term $\propto 1/v_F$ in the transverse cross-section (\ref{gopt}) was considered as anomalous and interpreted in the terms of spectral flow \cite{Vol93,VolB} though its original derivation from the partial-wave analysis\cite{Kop76} did not used this concept (see further discussion in Sec.~\ref{spFl}). The analysis presented here demonstrates that it can be explained within the framework of the scattering theory taking into account peculiarities of the Aharonov--Bohm effect for BCS quasiparticles.

\section{Bound vortex-core states and  the current in the core} \label{corSt}

\subsection{Bound Andreev states in a planar SNS junction} 

For the analysis of the role of the core states in vortex dynamics  it is useful  to consider a simplified approach to them based on geometric optics. Such an approach was suggested by \citet{stone96}. He also used the model of a normal vortex core exploiting its analogy with the 1D   problem of  Andreev bound states in the ballistic SNS junction. We also shall investigate this analogy.  In the past a number of authors addressed  the question whether and what Josephson current is possible through such a junction in full absence of the order parameter in the normal layer  \cite{Kul,Ishi,Bard}. They concluded that the Josephson current is possible due to phase coherence of the Andreev states, which are sensitive to the phase difference on the junction.

We consider a normal layer of the width $L$, which is perpendicular to the axis $y$. A superfluid  in superconducting regions $y<0$ and $y>L$ moves with the velocity $\bm v_s$. Let us look  for a state with the energy
\be
\epsilon =\epsilon_0 +\hbar \bm k \cdot \bm v_s \approx \epsilon_0 +\hbar \bm k_0 \cdot \bm v_s,
  \ee{spec}
where $|\epsilon_0|<\Delta$ and the wave vector $\bm k_0 (k_x, k_f, k_z)$ has a modulus equal to the Fermi wave number $k_F$, so that  the component $k_y$ is equal to $k_f=\sqrt{k_F^2-k_x^2-k_z^2}$.
The wave function, which satisfies the Bogolyubov--de Gennes equations,  is given by
\bem
\left(  \begin{array}{c}
u    \\
  v      
\end{array}  \right)
=\left(  \begin{array}{c}
A  e^{i m(\bm v_s\cdot \bm R )/\hbar +im\epsilon_0 y   /\hbar^2 k_f}   \\ B    e^{-i m(\bm v_s\cdot \bm R)/\hbar -im\epsilon_0 y   /\hbar^2 k_f} 
\end{array}  \right) e^{i \bm k_0\cdot \bm R} 
  \eem{norm}
 inside the normal layer $0<y<L$, 
 \bem
\left(  \begin{array}{c}
u    \\
  v      
\end{array}  \right)=\left(  \begin{array}{c}
u_- e^{i\theta_-/2+i m(\bm v_s\cdot \bm R )/\hbar }   \\
v_-    e^{-i\theta_-/2-i m(\bm v_s\cdot \bm R)/\hbar  }    
\end{array}  \right)
e^{i \bm k_0\cdot \bm R+{m\sqrt{\Delta^2-\epsilon_0^2}\over \hbar^2k_f}y }  
\nonumber \\        \eem{sup1}
inside the superconductor at $y<0$,  and
\bem
\left(  \begin{array}{c}
u    \\
  v      
\end{array}  \right)=C\left(  \begin{array}{c}
u_+ e^{i\theta_+/2+i m(\bm v_s\cdot \bm R )/\hbar }   \\
v_+    e^{-i\theta_+/2-i m(\bm v_s\cdot \bm R )/\hbar  }    
\end{array}  \right)
e^{i \bm k_0\cdot \bm R-{m\sqrt{\Delta^2-\epsilon_0^2}\over \hbar^2k_f}y }  
\nonumber \\
          \eem{sup2}
inside the superconductor at $y>L$. 
  Here 
\be 
u_\pm =v_\mp= \sqrt{{1\over 2} \left( 1\pm i  {\sqrt{\Delta^2-\epsilon_0^2}\over \epsilon_0}\right)},
  \ee{}
and the constants $A$ and $B$ are determined from the boundary conditions at the interface $y=0$: 
\be
A=u_- e^{i\theta_-/2},~~A=v_- e^{-i\theta_-/2}. 
  \ee{}
One can find the constant $C$ from the boundary conditions at the interface $y=L$ only for discrete values of the energy $\epsilon_0$ satisfying the following Bohr--Sommerfeld condition\cite{stone96}:
\be
\frac{2m \epsilon_0 L }{\hbar^2 k_f}=2\pi \left(s+{1\over 2}\right)+( \theta_+-\theta_-) -2\arcsin {\epsilon_0\over \Delta},
      \ee{}
with integer  $s$. At small energy $\epsilon_0 \ll \Delta$ this yields the spectrum of the Andreev bound states:
\be
\epsilon_0 =\left(\frac{2m L }{\hbar^2k_f}+{1\over \Delta}\right)^{-1}\left[2\pi  \left(s+{1\over 2}\right) +  (\theta_+-\theta_-) \right].
      \ee{}

The wave function given by Eqs.~(\ref{norm}) -  (\ref{sup2}) assumes that only the Andreev reflection occurs at the core boundaries, so the wave vector is always close to the Fermi surface and its normal component $k_y \approx k_f$  does not change a sign at the reflection. This is a valid assumption in the weak-coupling limit when  the superconducting gap $\Delta$ is small compared to the Fermi energy $\epsilon_F=\hbar^2 k_F^2/2m$.
In this approximation the boundary conditions at the interfaces $y=0$ and $y=L$ require continuity only of two spinor components, ignoring the continuity  conditions for spinor gradients. It is worthwhile of noting that this approximation does not provide exact continuity of the quasiparticle current at the interfaces as required by the the conservation law for quasiparticles. This can be achieved only in the next approximation taking into account the possibility of usual reflection changing the direction of the momentum normal to the layers. However, for our goals we need not the quasiparticle current but the genuine particle current, and the used approximation is sufficient.

\begin{figure*}
\includegraphics[width=0.85\textwidth]{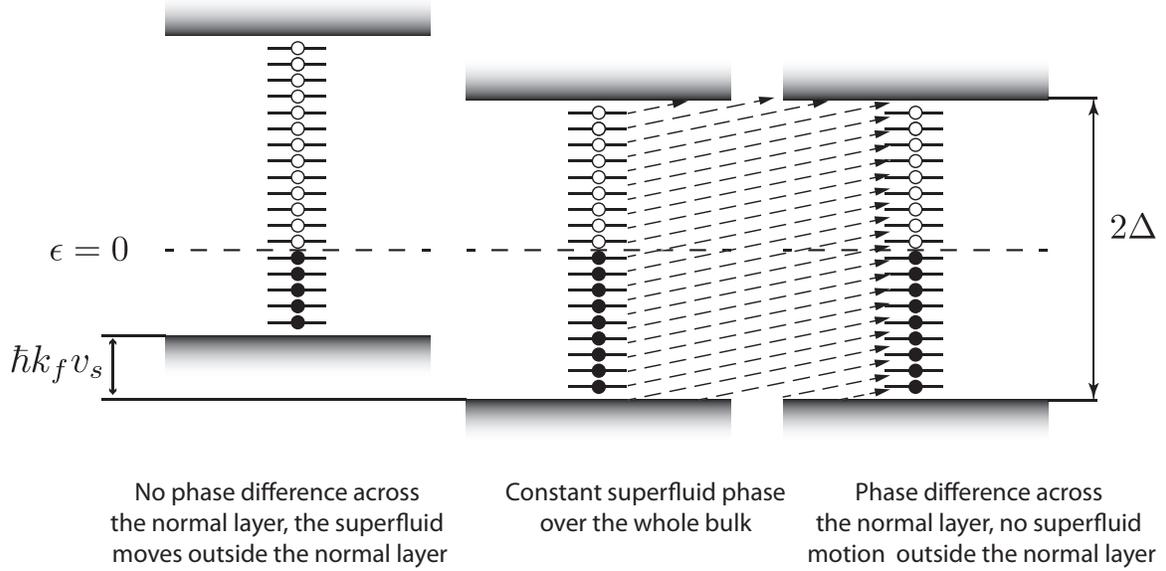}
\caption[]{Bound states inside the superconducting gap in the SNS junction. Occupied levels below the Fermi level $\epsilon =0$ are shown by black circles while empty levels above the Fermi level  are shown by white ones. {\em Center: } No phase difference across the normal layer, no superfluid velocity inside superconducting areas. {\em Right: } There is  phase difference $\theta_+-\theta_-$ across the normal layer but still no  superfluid velocity inside superconducting areas. Arrowed dashed lines show shift of level relatively to the gap and the Fermi level. Some levels exit from the gap at the top of the gap while some new levels enter the gap at the gap bottom. {\em Left: } No phase difference across the normal layer, but there is the superfluid velocity $v_s$ at the bulk of superconductors. This  shifts the gap with respect to the Fermi level and changes the number of occupied levels.  }
\label{Spec}
\end{figure*}

 Now let us find  the contribution of bound states to the momentum of the liquid. 
Since  in the bound states particle and hole components are equal in amplitude,    the charge $N(\bm k) =|u|^2-|v|^2$ in these states vanishes, and according  to \eq{cur1} the current normal to the layers  in any bound state is about $\hbar   k_f/m$. Let us consider the case of a wide normal layer $L \gg \hbar k_F/m\Delta$ when there is a large number of bound states and the sum over them can be replaced by an integral. Then the total contribution of the bound states to the current is simply a product of $\hbar   k_f/m$ and a difference of the number of states with the wave vectors in opposite directions. At $T=0$ all states with $\epsilon =\epsilon_0+\hbar \bm k_0 \cdot \bm v_s <0$ are filled. This yields that the total momentum in the bound states (per unit area of the SNS junction) normal to the layers is
\be
P_{bs}=- \int \limits_{k<k_F}{ \hbar^2 k_f^2  v_s\over  \delta \epsilon}{d\bm k_\parallel\over 4\pi^2} = -L n  m v_s,
  \ee{Pn} 
where $\bm k_\parallel(k_x,k_z)$ is the component of the wave vector in the layer plane,  $n$ is the particle density and
\be
 \delta \epsilon  =\pi \frac{\hbar^2k_f}{m L }
      \ee{}
 is the distance between discrete levels. The momentum corresponds to the current density $j_{bs}=P_{bs}/Lm =-n v_s$ inside the normal layer. This shows that quasiparticles in core bound states play a role of the effective normal component, which exists even at zero temperature \cite{VolB}.

 Note that the phase difference $\theta_+-\theta_-$ between the superconducting banks of the SNS junction has no effect on the current in the continuum limit. This is because the main effect of the phase difference  is a shift of the levels in the forbidden gap without any essential change of their number. The shift of the levels leads to an entry of a new level on one side of the gap and an exit of an old level on another side (Fig.~\ref{Spec}). Normally these two events are not fully synchronized and the number of level can fluctuate with $\pm 1$ level. This fluctuation is not essential in the limit of large number of levels.  On the other hand, exactly this small variation of the level number leads to the Josephson effect in the SNS junction \cite{Kul,Ishi,Bard}, which is beyond the scope of the  present work.
 The shift of levels constitutes the phenomenon of spectral flow, which arises if the phase difference 
 $\theta_+-\theta_-$ monotonously increases or decreases and the levels cross the forbidden gap. 
  
 One should also consider the contribution of the continuum delocalized states with negative energy, which are fully occupied in the ground state. 
For a delocalized state ($|\epsilon_0| >\Delta$)  of a quasiparticle propagating from $y=-\infty$ to $y=\infty$, the spinor  in the normal layer $0<y<L$ is given by the same expression as \eq{norm} for the bound state, whereas in superconducting layers the states are described by spinors 
\bem
t\left(  \begin{array}{c}
u_0 e^{i\theta_+/2+i m(\bm v_s\cdot \bm R )/\hbar }    \\
 v_0 e^{-i\theta_+/2-i m(\bm v_s\cdot \bm R)/\hbar }          
\end{array}  \right)
e^{i \bm k_0\cdot \bm R+i {m\sqrt{\epsilon_0^2-\Delta^2}\over \hbar^2k_f}  y} 
      \eem{}
for $y>L$ and 
\bem
e^{i \bm k_0\cdot \bm R }\left[\left(  \begin{array}{c}
u_0 e^{i\theta_-/2+i m(\bm v_s\cdot \bm R )/\hbar }    \\
 v_0 e^{-i\theta_-/2-i m(\bm v_s\cdot \bm R )/\hbar }          
\end{array}  \right)e^{i {m\sqrt{\epsilon_0^2-\Delta^2}\over \hbar^2k_f}  y} 
\right. \nonumber \\ \left. +r\left(  \begin{array}{c}
v_0 e^{i\theta_-/2+i m(\bm v_s\cdot \bm R )/\hbar }    \\
 u_0 e^{-i\theta_-/2-i m(\bm v_s\cdot \bm R)/\hbar }          
\end{array}  \right)e^{-i {m\sqrt{\epsilon_0^2-\Delta^2}\over \hbar^2k_f}  y}  \right] . \nonumber \\
      \eem{}
for $y<0$. Here $t$ and $r$ are amplitudes of transmission and reflection ($|t|^2+|r|^2=1$) which are determined from the continuity of spinor components (but not their derivatives!)\cite{Bard} at $y=0$ and $y=L$. As well as for bound states, the analysis considers only the Andreev reflection neglecting  probability of usual reflection, which changes the direction of the wave vector. 
The amplitudes of the spinor components in the normal layer [see \eq{norm}] are 
\bem
A=t u_0e^{i\theta_+/2+im(\sqrt{\epsilon_0^2-\Delta^2}-\epsilon_0)L/\hbar^2k_f },
\nonumber \\
B=tv_0e^{-i\theta_+/2+im(\sqrt{\epsilon_0^2-\Delta^2}+\epsilon_0)L/\hbar^2k_f }.
    \eem{} 
The transmission probability is 
\be
|t|^2= \frac{\epsilon_0^2-\Delta^2} { \epsilon_0^2-\Delta^2 \cos^2[\epsilon_0 m L /\hbar^2 k_f-(\theta_+-\theta_-)/2]  } . \ee{}
The transmission probability differs from unity in the small energy interval of the order $\Delta \ll \epsilon_F$, and the effect of reflection is not essential for the contribution of delocalized states to the supercurrent. The latter can be found by summation of the Eq.~(\ref{cur1}) over the whole continuum of free bulk states. The whole particle density is accumulated in delocalized but not bound states.  Neglecting reflection for the continuum states, the density and the current in the normal and the superconducting areas do not differ essentially. So the whole ensemble of delocalized quasiparticles is a  liquid of nearly constant density $n$ moving with the spatially uniform velocity $v_s$. This points out nearly ideal transparency of the ballistic normal layers for the supercurrent of delocalized quasiparticles. Note that scattering of continuum states by impurities is impossible since all continuum states are fully occupied.

Summing the momenta in localized and delocalized states inside the normal layer, one obtains that the total momentum and the current eventually vanish there (with accuracy of the small parameter of weak coupling  $\Delta/\epsilon_F$). Keeping in mind the presence of the current $nv_s$ in superconducting layers, this violates the conservation law for the particle number, since backflow  in our 1D geometry is impossible and the current must be constant along the direction normal to the layers.  A proper conclusion from this is that the superfluid transport (but not diffusive transport with dissipation!) with high superfluid velocity in this one-dimensional geometry is impossible. But this  does not rules out the superfluid transport with very low superfluid velocities $v_s \leq \hbar /mL$ when discreteness of the Andreev  bound states and the phase difference across the normal layer cannot be ignored. This returns us again to the problem of the Josephson effect in the SNS junction \cite{Kul,Ishi,Bard}.

\subsection{Bound vortex-core states: ballistic  normal core} \label{bnc}

\begin{figure}
\includegraphics[width=.35\textwidth]{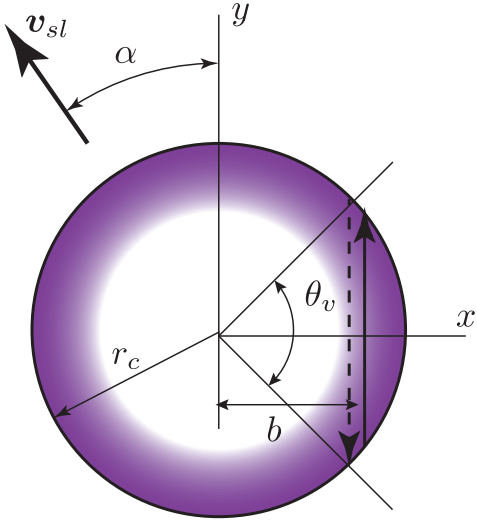}
\caption[]{(Color online) Bound state in the normal core. The vertical solid arrowed line shows the trajectory of the quasiparticle and the vertical dashed arrowed line shows the trajectory of the quasihole after Andreev reflection of the quasiparticle at the core boundary. Note that the picture is purely schematic, and in fact the analysis was done for the case when the impact parameter $b$ is much less than the core radius $r_c$ but still much larger than the interatomic distance $~1/k_F$.}
\label{norC}
\end{figure}

Now let us consider bound states  in a normal core of a vortex. 
A reliable assumption is that a quasiparticle inside the core, where the superconducting order parameter vanishes,  moves along an approximately straight trajectory  back and forth reversing  its direction of motion via Andreev reflection at the boundary of the core. The trajectory is chosen to be parallel to the $y$ axis.   For trajectories with impact parameters much less than the core radius the  bound states are similar to those in the SNS junction with the normal-layer width $L$ equal to the core diameter $2r_c$. On the other hand the phase difference   $\theta_+-\theta_-=\theta_v+\theta_s$ consists from the phase difference induced by the vortex, $\theta_v=\pi- 2 \arcsin (b/r_c)\approx \pi- 2 b/r_c$, and the phase difference $ \theta_s $ produced by the superflow past the vortex. 
Here $b= l /k_f$ is the impact parameter and $l$ is the quantum number of the discrete angular momentum. 
Geometry of the process is shown in Fig.~\ref{norC}.

Eventually the energy   of the bound state in the normal core for the chiral zero-crossing branch $s=-1$ depends on the orbital quantum number $l$ and is $\epsilon (l)= \epsilon_0(l) + \hbar k_f v_{sl} \cos \alpha $, where 
\be
\epsilon_0(l) =\left(1+\frac{\hbar^2k_f}{2m r_c  \Delta}\right)^{-1}\frac{\hbar^2k_f}{2m r_c }\left(-{b \over r_c}+ {\theta_s\over 2} \right), 
      \ee{spG}
and $\alpha$ is the angle between  the trajectory (the axis $y$) and the local  superfluid velocity $\bm v_{sl}$ (see Fig.~\ref{norC}), which is different from the superfluid velocity $\bm v_s$ far from the vortex in the presence of the backflow (see below).
Introducing the angular momentum $L_z=\hbar l $ of the bound state, there is a  frequency
\be
\omega_0 ={\partial \epsilon_0\over \partial L_z}=  {  \hbar  \over 2mr_c^2 } \left(1+\frac{\hbar^2k_f}{2m \Delta r_c }\right)^{-1},
                 \ee{omega}
with which the trajectory slowly  rotates around the vortex axis. The phase difference  from the superflow outside the core can be presented in the form of a dipole field $\theta_s=(1+\hbar^2k_f/2m r_c  \Delta)(4m/\hbar) \bm r_c  \cdot \bm v_\theta$, where $\bm r_c$ is the vector of the modulus $r_c$ directed normally to the cylindric border of the core and the superfluid velocity $\bm v_{\theta}$ is one more superfluid velocity different in general from the asymptotic velocity $\bm v_s$ and the local velocity $\bm v_{sl}$. Introducing 
the isotropic part of the spectrum 
\be
     \epsilon_{00}= - \omega_0 L_z
                 \ee{en}
the whole spectrum becomes
\be
\epsilon =\epsilon_{00} +\hbar \bm k \cdot (\tilde{\bm  v}_s  - \bm v_L),
     \ee{}
where the presence of the vortex velocity $\bm v_L$  points out that the calculation is done for the coordinate frame connected with vortex and 
\be
 \tilde {\bm v}_s =\bm v_{\theta}  +\bm v_{sl} 
     \ee{efV}
is the effective superfluid velocity taking into account the superfluid velocity $\bm v_{sl} $ outside the core and the contribution of the phase difference $\theta_s$. All three velocities in \eq{efV} are parallel to each other.

For the analysis of the vortex mass one needs to know the contribution of bound states  to the total momentum of the liquid if the fluid flows past the vortex with the superfluid velocity $\bm v_s$. As well as in the case of the SNS junction, every bound state has a momentum of the magnitude about $\hbar k_f$ directed along the bound-state trajectory. 
 Taking into account that the energy interval between levels is $\delta \epsilon =\hbar \omega_0$ and integrating over all directions and the wave number component $k_z$ along the vortex axis one obtains that the total momentum is 
\bem
\bm P_{bs}=- {1\over 2}\int \limits_{-k_F}^{k_F}{dk_z\over 2\pi} \int d\alpha\,\cos^2\alpha{\hbar k_f ^2  \over \omega_0}(\bm v_{sl}-  \bm v_L)
\nonumber \\
 = -{\pi \hbar n \over \omega_0}(\bm v_{sl}-  \bm v_L).
  \eem{Pbs} 
The expression was derived for large $r_c$ neglecting corrections of the order  $\hbar^2k_f/2m \Delta r_c $ in \eq{omega}. The coefficient before the velocity $\bm v_L$ is the Kopnin vortex mass $\mu_K=\pi \hbar n/\omega_0$. Using \eq{omega} for $\omega_0$ in the limit of large core $r_c \gg \hbar^2 k_F /m\Delta$ and assuming that the momentum is uniformly distributed over the area $\pi r_c^2$ of  the core this momentum corresponds to the current $\bm j_{bs}=\bm P_{bs}/\pi r_c^2m =- 2n(\bm v_{sl}-  \bm v_L)$ inside the core.

The momentum in the core depends only on the local superfluid velocity $\bm v_{sl}$ outside the core and not on the phase difference $\theta_s$. The reason is the same as in the case of the SNS junction: the phase difference shifts energy levels but does not change their number since any crossing  of the zero energy by a level is compensated by an entry or an exit of an level at the bottom of the forbidden gap. 
Meanwhile, it is variation of  the phase difference  that governs the spectral flow.  In contrast to the SNS case, 
where the phase difference across the normal layer can vary monotonously, the phase difference across the normal core of the moving vortex can only oscillate  without monotonously crossing the forbidden gap. This rules out the steady spectral flow. The oscillation is related with rotation of the bound state with angular velocity $\omega_0$ and dependence of the level position with respect to the gap on the $\alpha$-dependent phase difference $\theta_s$ in \eq{spG}.\cite{stone96}

In the model of the normal core the Andreev reflection for all bound states occurs at the core boundary, and the energy of bound states are easily determined analytically from the semiclassical Bohr--Sommerfeld condition. Meanwhile, the more realistic model of the core with linear growth of the order parameter  in the core considered in App.~\ref{bss} shows that though the concept of well defined trajectory (geometric optics) works well, one cannot use the semiclassical approach for description of motion along the trajectory and the Bohr--Sommerfeld condition is invalid. Despite this, the model of the normal core gives a qualitatively correct energy spectrum, different from that from more accurate theories only by a numerical factor. On the other hand, this model allows a simple analytical analysis of the backflow effect on the vortex mass, which would require less transparent numerical calculations in more realistic models.

\section{Vortex mass in the Bose liquid: backflow and compressibility contributions} \label{vmB}

  In an ideal liquid a singular vortex   line has no own inertia and cannot move with respect to the liquid, in which the vortex line is immersed (Helmholtz's theorem). But this statement is exact only in the limit of an infinitely thin line. Taking into account the finite size of the vortex core the vortex line can move with its own velocity $\bm  v_L$ different from that of the surrounding  
liquid and there is an inertial force proportional  to the vortex line acceleration  $d\bm  v_L/dt$.

  A naive estimation for the vortex mass is to deduce it from the picture of a cylinder without own mass moving through a perfect fluid assuming that the cylinder has a radius equal to the core radius. \cite{Bay83} Then classic hydrodynamics tells that the cylinder induces a dipole velocity field around it (backflow): 
\begin{equation}
{\bm V}_{ bf}(\bm {r})={\kappa\over 2\pi}\bm \nabla \theta_{bf} = -r_c^2\bm  \nabla \left[ \frac{{\bm v}_{bf} \cdot {\bm r}} { r^2}\right].
\label{bf}
\end{equation}
The condition of the absence of the radial current through the core (cylinder) boundary in the coordinate frame moving the vortex velocity $\bm v_L$ requires that the constant velocity ${\bm v}_{bf}$, which determines the backflow, is  ${\bm v}_{bf}=\bm v_L-\bm v_s $. Here $\bm v_s$ is the superfluid velocity far from the vortex core. We consider the case of $T=0$ when $n_s=n$. The kinetic energy of the backflow is given by
\begin{eqnarray}
\mu_v\frac{({\bm v}_L - {\bm v}_s)^2}{2} 
= \frac{mn r_c^4}{2} \int \limits_{r>r_c} d\bm r ^2\, \left|\nabla 
\left[\frac{({\bm v}_L- {\bm v}_s) \cdot {\bm r}}{r^2}\right]
\right|^2 
\nonumber \\
= \pi r_c^2 mn \frac{({\bm v}_L- {\bm v}_s)^2}{2}.~~~~
\label{bfM}
\end{eqnarray}
So this yields the vortex mass $\mu_v$ equal to   $\mu_{core} = \pi r_c^2 mn $, which is a mass per unit length of the liquid inside a cylinder of the radius equal to the core radius $r_c$\cite{Bay83}. Later we shall call $\mu_{core}$ a core mass (in contrast to a more general term vortex mass taking into account all possible contributions to the mass of the vortex).

The vortex mass can be determined from calculation of the vortex-velocity dependent contribution to the energy or the momentum. Naturally the both calculations should yield the same mass. Sometimes it is simpler to calculate  the momentum \cite{Kop}. But calculation of the momentum of the backflow has a subtlety well known in classical hydrodynamics. The direct way to estimate the momentum of the potential velocity field in an incompressible liquid is to integrate the expression for the momentum by parts. For the backflow this yields
\bem
\bm P =mn {\kappa\over 2\pi}\int \bm \nabla \theta_{bf}(\bm r) \,d\bm r
\nonumber \\
=mn{\kappa\over 2\pi}\left( \int \limits_{r=r_c} \theta_{bf} \bm n dS-\int\limits_{r\to \infty} \theta_{bf} \bm n dS\right),
 \eem{}
where integration is reduced to integrals over the cylindric surfaces of radius $r_c$ and of infinite radius and $\bm n$ is a normal to these surfaces. Strictly mathematically speaking this yields zero since surface integrals do not depend on surface radii for the backflow field. However, one should take into account that any finite momentum $\bm P$ in an incompressible liquid means that the whole liquid moves with the velocity $\bm P /mn V$ inversely proportional to the volume $V$. One should take into account this tiny velocity simply by deleting the contribution from the distant surface. This yields the momentum called in hydrodynamics Kelvin impulse (see Sect.~119 in the textbook by \citet{Lam}):
\bem
\bm P_K=mn{\kappa\over 2\pi} \int \limits_{r=r_c} \theta_{bf} \bm n dS= \mu_{core} \bm v_{bf}.
 \eem{Kel}
In classical hydrodynamics they  justify using the Kelvin impulse for an object moving through an incompressible liquids by considering the momentum transferred to the object when making it to move from rest.\cite{Lam} But in quantum hydrodynamics the justification looks simpler. Local perturbations of the velocity field cannot change the phase at infinity. So the boundary condition at infinity is not vanishing velocity, but vanishing phase, i.e., the potential of the velocity field. On the basis of it the integral over the distant surface in the expression for the momentum should be ignored. 

The calculation of the vortex mass  assumed that a moving core is  impenetrable for the fluid as a real rigid cylinder though the cylinder itself has no mass. In reality the vortex core is not empty and   the superfluid  will flow through the core, thus producing a reduced backflow field.\cite{Son98} So our simple calculation provides only an upper bound on the vortex mass related to the core. For illustration of this effect let us consider the model of a partially filled core with constant particle density $n(1-\lambda)$ inside characterized by the parameter $\lambda<1$. Inside the core the liquid moves with the constant velocity $\bm v_{in}$, which corresponds to the phase $\theta_{in} =2\pi (\bm v_{in}\cdot \bm r)/\kappa$. The continuity of the phases $\theta_{in}$ inside the core and the phase $\theta_{out}=2\pi ((\bm v_s-\bm v_L)\cdot \bm r)/\kappa+\theta_{bf}$ outside the core together with continuity of the radial flow at the core boundary yield:
\be
\bm v_{in}=\bm v_s-\bm v_L-\bm v_{bf}, ~~\bm v_{bf} =-{\lambda\over 2-\lambda}(\bm v_s-\bm v_L).
      \ee{}
In the coordinate frame moving with the vortex velocity this gives the momentum
\bem
\bm P_L= \pi r_c^2 mn (1-\lambda) \bm v_{in} + (S- \pi r_c^2) nm({\bm v}_s - {\bm v}_L) 
\nonumber \\
+P_K=nm(\bm v_s -\bm v_L)\left(S-\pi r_c^2   {2\lambda\over 2-\lambda} \right),
  \eem{}
where $S$ is the whole area occupied by the liquid. In order to see the value of the vortex mass one needs to know the momentum in the arbitrary coordinate frame:
\bem
\bm P= \bm P_L+ mn[ S-\pi r_c^2+ (1-\lambda) \pi r_c^2]\bm v_L
\nonumber \\
=mn \left(S-\pi r_c^2   {2\lambda\over 2-\lambda} \right)\bm v_s+mn\pi r_c^2   {\lambda^2\over 2-\lambda}\bm v_L .
  \eem{}
The vortex mass $\mu_v=\mu_{core}\lambda^2/ (2-\lambda) $ is  a factor before the vortex velocity $\bm v_L$. If the density suppression $\Delta n=n\lambda$ in the core is small the vortex mass is quadratic in $\Delta n$. 

Strictly speaking the model of constant density in the core is not relevant for singular vortices in  Bose superfluids where the density must vanish on a vortex axis. 
Therefore, in App.~\ref{bm}  we derive the vortex mass for the Bose superfluid using a more realistic model with linear in $r$ growth of the density in the core. 
On the other hand, the model of constant density in the core is relevant for continuous vortices in the Fermi liquids, namely for estimation of the effect of superfluid density suppression on the vortex mass. However, this contribution is small compared to the effect of bound states in the core (see Sec.~\ref{fermiM}).  

But in the Bose liquid the most important contribution to the vortex mass is connected with finite compressibility of the liquid. The cross term in the kinetic energy of the velocity field $ \bm  v_s(\bm  r)-\bm  v_L=\bm  v_v(\bm  r)+\bm  v_s-\bm  v_L$ in the coordinate frame moving with vortex produces the  density variation in accordance with the Bernoulli law:
\be
\delta n = -mn{\partial n \over \partial P} \bm  v_v(\bm  r)\cdot(\bm  v_s-\bm  v_L) =-{n \over c_s^2}\bm  v_v(\bm  r)\cdot(\bm  v_s-\bm  v_L), 
       \ee{bern}
where $\partial n/ \partial P=1/mc_s^2$ is the fluid compressibility, $P$ is the pressure, and $c_s$ is the sound velocity.  The density variation leads to 
the energy contribution  \citep{Duan93,Duan}
\begin{eqnarray}
\mu_{com}\frac{({\bm v}_L-{\bm v}_s)^2}{2} = \int \limits_{r>r_c} d\bm r^2 \, \frac{\partial^2 E} 
{\partial n^2} \frac{\delta n^2}{2} 
\nonumber \\ 
= \int \limits_{r>r_c} d\bm r^2 \, \frac{\partial \mu} 
{\partial n } \frac{\delta n^2}{2} = \frac{\varepsilon}{ c_s^2}
\frac{({\bm v}_L - {\bm v}_s)^2}{2},\label{compM}
\end{eqnarray}
where $\mu=\partial E/\partial n$ is the chemical potential, $\partial \mu/\partial n= mc_s^2/n$, and 
\[
 \varepsilon={mn \kappa^2\over 4\pi}\ln \frac{R}{r_c}
 \]
  is  the static vortex energy  per unit vortex-line length. Like the vortex energy, the vortex mass is determined by  a logarithmically divergent integral, which has to be cut off at some hydrodynamic scale $R$, e.~g., the intervortex distance. In the Bose superfluid, according to the Gross-Pitaevskii theory,  the core radius $r_c \sim \kappa/c_s$ is also determined by the sound velocity $c_s$ and as a consequence, the compressibility mass is by the logarithmic factor larger than the core mass $\mu_{core} = \pi r_c^2 mn$. 

\section{Vortex mass in the Fermi superfluid} \label{fermiM}

The two contributions to the vortex mass (from the backflow and the liquid compressibility) in the Bose superfluid  in principle are relevant also in the Fermi superfluid. However, the compressibility becomes inessential in the weak-coupling limit despite a large logarithmic factor. The difference with the Bose superfluid is that while in the Bose superfluid the sound velocity goes down (compressibility goes up)  in the weak-interaction limit, in the Fermi superfluid the sound velocity remains high being always of the order of the Fermi velocity.
But the most important difference between the Bose and the Fermi superfluids comes from bound core states, which
 contribute not only to the mutual friction force but also to the vortex mass \cite{Kop}. 
 Earlier we derived the momentum $\bm P_{bs}$ in the ground state in the presence of the superflow past the vortex [\eq{Pbs}]. 
  The factor before the vortex velocity $\bm v_L$  in this expression is the Kopnin mass  $\mu_K=\pi \hbar n/\omega_0$.   
However the full vortex mass is not reduced to the Kopnin mass.  The current $\bm j_{bs}=\bm P_{bs}/\pi mr_c^2$  in the bound states exists only inside the core and must transform to the superfluid current outside the core. The latter current forms the backflow velocity field, which must be determined from the continuity equation for the total fluid. As a result, the Kopnin mass  will be renormalized by the backflow effect. 

In analogy with the analysis of the backflow for the Bose liquid, the local superfluid velocity ${\bm v}_{sl}=\bm v_s+\bm v_{bf}$ at the core boundary  differs from  the superfluid velocity far from the vortex and the continuity of the current at the core boundary is
\bem
\bm j_{bs} ={\hbar \over \omega_0 mr_c^2} n(\bm v_L -{\bm v}_s-\bm v_{bf})
= n \bm v_{bf}. 
     \eem{}
Note that the current in the continuum of delocalized states does not affect this condition because it has no discontinuity at the core boundary and contributes the same term $mn \bm v_{sl}$ to the two sides of this equation. The latter yields 
\bem
\bm v_{bf}=\frac{\mu_K}{\mu_{core}+\mu_K}(\bm v_L -\bm v_s),
  \eem{}
and the total momentum including the backflow momentum (Kelvin impulse) $\bm P_K$ given by \eq{Kel} is
\bem
\bm P_{bs} +\mu_{core} \bm v_{bf}=\frac{2\mu_{core}\mu_K}{\mu_{core}+\mu_K}(\bm v_L -{\bm v}_s).
  \eem{}
According to this expression the Kopnin mass $\mu_K$ is renormalized by the factor $2\mu_{core} /( \mu_K+\mu_{core})$ equal to 4/3 for the value of $\omega_0$ given by \eq{omega}  in the limit of large core radius $r_c\gg \hbar^2k_f/2m \Delta $. The most important outcome of this analysis is not this numerical factor, which depends on the model of the core anyway, but a more adequate insight into the origin of the vortex mass. If the Kopnin mass $\mu_K$ is much smaller than the core mass $\mu_{core}$, the Kopnin mass is renormalized by the factor 2, i.e., the backflow gives the same contribution as the bare Kopnin mass. The case of small normal density of bound states is realized for a vortex with a continuous core in superfluid $^3$He when the core radius $r_c$ essentially exceeds the coherence length $\xi_c =\hbar v_F/\Delta $ and  $\mu_K   \sim \mu_{core} \xi_c /r_c$. Addressing this case, \citet{Vol98,VolB} arrived to an incorrect conclusion that the contribution of the backflow to the vortex mass is negligible compared to the bare Kopnin mass. The reason for it was that Volovik used the condition of continuity of the superfluid component (see Eq.~(24.16) in his book\cite{VolB}),  whereas only the total particle number of the liquid       but not  its superfluid part is conserved in the presence of the Andreev reflection. In fact, Volovik estimated the backflow effect from weak suppression of the superfluid density inside the continuous core considered for the Bose liquid in the previous section. He ignored the backflow induced by the current in bound states.

\section{Boltzmann equation for the core-states quasiparticles: the Kopnin--Kravtsov force and the vortex mass} \label{KE}

If there are impurities in superconductors or collisions of bound quasiparticles with free bulk quasiparticles in superfluid $^3$He, the bound states produce not only the vortex mass but also  the  mutual friction force (Kopnin--Kravtsov force). In this case one should use  the Boltzmann equation.\cite{Kop}  Let us write the  Boltzmann equation in the continuum of semiclassical states bound in the core and characterized by the two Hamiltonian-conjugate quantities ``angle $\alpha$ - moment $L_z$'':
\bem
{\partial f\over \partial t}- {\partial\epsilon \over \partial \alpha }{\partial f\over \partial L_z}+{\partial\epsilon \over \partial L_z}{\partial f\over \partial \alpha}=\left. {\partial f\over \partial t} \right|_{col}.
        \eem{kinO}
The collision term in the right-hand side in the relaxation-time approximation is 
\be
\left. {\partial f\over \partial t} \right|_{col}=-{f- f_n(\epsilon, \bm v_n)\over \tau}.
    \ee{}
It takes into account elastic collisions with impurities in superconductors (then $\bm v_n$ is the velocity of the crystal lattice) or with bulk free quasiparticles in superfluids. Here
\be 
f_n(\epsilon, \bm v_n)= \frac{1}{e^{\epsilon -\hbar \bm k \cdot (\bm v_n-\bm v_L)\over T}+1}
= \frac{1}{e^{\epsilon_{00} -\hbar \bm k \cdot (\bm v_n- \tilde {\bm v}_s)\over T}+1}
   \ee{fn}
is the distribution function for bound states, which are in the equilibrium with the normal component. 

The equilibrium distribution function in the collision term has a small anisotropic part if the superfluid part moves with respect to the normal part of the liquid. This is well known property of the Boltzmann equation in superconductors \cite{Aro, stone96,VolB}. Note that \citet{Kop} used the different Boltzmann equation, which follows from that used in the paper if the superfluid velocity $\tilde {\bm v}_s$ is replaced by the normal velocity $\bm v_n =0$. This difference does not lead to the difference in the Kopnin--Kravtsov force and the Kopnin mass, since they do not depend on the relative velocity $\tilde {\bm v}_s- \bm v_n$,. But in general it could be important, e.g., for non-stationary phenomena when the distribution function varies in time.

We expand the distribution functions around the isotropic equilibrium distribution function  $f_0(\epsilon_{00})$:
\bem
f(\bm p)=f_0 (\epsilon_{00}) +f_1(\epsilon, \bm v_n),
    \eem{}
The zero-approximation function $f_0(\epsilon_{00})$ is the equilibrium Fermi distribution function equal to $f_n$  at $ \bm v_n=\bm v_s =\bm v_L$.  The equation for the first-order correction linear in  the relative velocities is
\bem
\hbar \omega_0 \bm k \cdot [( \tilde {\bm v}_s-\bm v_ L)\times \hat z] {\partial f_0\over \partial \epsilon }-\omega_0{\partial f_1\over \partial \alpha}
\nonumber \\
=-{1\over  \tau}\left[f_1 -\hbar \bm k\cdot (\bm v_n- \tilde {\bm v}_s){\partial f_0\over \partial \epsilon}\right].
    \eem{kin}
Its solution is
\bem 
f_1 = {\partial f_0\over \partial \epsilon}\hbar\left[\bm k\cdot ( \tilde {\bm v}_s-\bm v_L)
\right. \nonumber \\ \left.
 -\frac{\omega_0 \tau\bm k \cdot [\bm (\bm v_n-\bm v_L)\times \hat z]+\bm k\cdot (\bm v_n- \bm v_L)}{1+\omega_0^2\tau^2} \right].
       \eem{f1}

The total momentum in the vortex-core bound states\cite{KopM,Kop} is given by
\bem
\bm P_{bs}=  {1\over 2}\int \limits_{-k_F}^{k_F}{dk_z\over 2\pi}  \int d\alpha
 \int\limits_{L_z^{min}} ^{L_z^{max}}{dL_z \over 2\pi}\bm k f(\alpha, L_z).
   \eem{Pkin}   
Here $L_z^{max} =L_0+ \hbar k_f  r_c \theta_s/2$ and $L_z^{min} =-L_0+ \hbar k_f  r_c \theta_s/2$ are maximal and minimal values of the angular momentum in the bound state, which differ from $\pm L_0=\pm \Delta /\omega_0$ because of the phase shift $\theta_s$. This may look as if the momentum  depends on the phase shift $\theta_s$ contrary to what was calculated for the ground state in Sec.~\ref{bnc}. Indeed, the anisotropic part of the distribution function $f_1$ obtained from the Boltzmann equation  depends on the $\theta_s$-dependent $\tilde v_s$  given by \eq{efV}. This is a natural result since the Boltzmann equation takes into account only events near the Fermi surface, while entries and  exits of the bound states to and from the forbidden gap at the top and at the bottom of the gap are also important, as was demonstrated above. In fact these events are accounted for with direction-dependent  limits $L_z^{max} $ and $L_z^{min} $ of the integral in \eq{Pkin}. One may change variable in this integral introducing the modified angular momentum   $L'_z=L_z-  \hbar k_f  r_c \theta_s/2$ so that 
\bem
\bm P_{bs}=\int \limits_{-k_F}^{k_F}{dk_z\over 4\pi}\int d\alpha \int\limits_{-L_0  } ^{L_0 }{dL'_z \over 2\pi}\bm k  
\left[f_1-{\partial f_0\over \partial \epsilon}\hbar\bm k\cdot ( \bm v_\theta-\bm v_L)\right]. 
\nonumber \\
   \eem{}   
The second term in brackets cancel the $\theta_s$ dependent term in $f_1$, and eventually after using the distribution function given by \eq{f1}  only the local superfluid velocity $\bm v_{sl}$ appears in the  expression for the total momentum of core bound states. One can use the modified angular momentum $L'_z$ as a new variable instead of $L_z$ from the very beginning in the Boltzmann equation  (\ref{kinO}) itself with the same result: the phase difference $\theta_s$ drops out from all expressions and the effective superfluid velocity $\tilde {\bm v}_s$ reduces to the local superfluid velocity $\bm v_{sl}$ outside the core.

In the limit of zero temperature $\partial f_0/ \partial \epsilon=-\delta(\epsilon)$ and the momentum in the bound states is
   \bem
\bm P_{bs} ={\pi \hbar n\over \omega_0}\left\{\bm v_L -{\bm v}_{sl}+\frac{\omega_0 \tau [\bm (\bm v_n-\bm v_L)\times \hat z]+ \bm v_n-\bm v_L}{1+\omega_0^2\tau^2} \right\}. 
\nonumber \\
   \eem{PbsB}   
The expression reduces to \eq{Pbs} in the limit $\tau \to \infty$.
The part of the momentum linear in $\bm v_L$ determines the Kopnin mass taking into account the effect of collisions. The mass becomes a tensor:
\be
\hat \mu_K =\frac{\pi \hbar n \tau}{1+\omega_0^2\tau^2} \left(  \begin{array}{cc}
 \omega_0 \tau  &   -1      \\
1 &     \omega_0 \tau            
\end{array}  \right).
  \ee{}
\citet{KopV}  called the term  in the momentum transverse to the relative velocity $\bm v_n-\bm v_L$  {\em transverse vortex mass}. This term, however, does not lead to a conservative inertial force, which follows from some Hamiltonian. It determines a high-frequency  correction to the dissipative (longitudinal) mutual-friction force, which has its counterpart in the dissipative function (see the next section).

Repeating the process of renormalization of the Kopnin mass by the backflow effect, one obtains the same renormalization  factor $2\mu_{core} /( \mu_K+\mu_{core})$ as obtained in the previous section without collisions.  In  this factor the Kopnin mass   $ \mu_K=\pi \hbar n/\omega_0$ is the scalar mass in the limit $\tau \to \infty$.

In the case of frequent collisions ($\tau \ll 1/\omega_0$) the velocity $\bm v_L$ drops out from the expression (\ref{PbsB}) for the momentum, and the Kopnin mass vanishes. This is because in this limit the effect of reflections from the walls of the core is fully suppressed by frequent collisions with impurities or quasiparticles. This does not mean the total absence of the vortex mass but its absence {\em in our approximation}, which neglected effects of the order $\Delta/\epsilon_F$. It is worthwhile to note that the small $\omega_0\tau$ does not necessarily invalidate the assumption  that the mean-free path $l_f$ of quasiparticles is much longer than the core radius mentioned in Introduction. Indeed, since $\tau =l_f/v_F$ and $\omega_0\sim \hbar/mr_c^2$ the condition $\omega_0\tau\ll 1$ reduces to the condition $l_f/r_c \ll \epsilon_F/\Delta$. In the weak-coupling limit $\epsilon_F/\Delta$ is very large so even large $l_f/r_c$ can satisfy this condition.

The contribution of the bound states to the mutual-friction force [see \eq{vd}] is determined by the momentum transferred from bound states confined in the vortex core to normal quasiparticles or impurities via collisions:
\bem
\bm F_c = {1\over 2}\int \limits_{-k_F}^{k_F}{dk_z\over 2\pi} \int d\alpha
 \int\limits_{L_z^{min}} ^{L_z^{max}}{dL_z \over 2\pi}\bm k \left. {\partial f\over \partial t} \right|_{col}.
   \eem{colInt}   
Substituting $f_1$ from \eq{f1} the core contribution to the mutual-friction force is  
\bem
\bm F_c=\pi  \hbar n \frac{\omega_0 \tau   (\bm v_n-\bm v_ L) - [(\bm v_n-\bm v_ L)\times \hat z] }{1+\omega_0^2\tau^2}.
      \eem{fc}
The force component transverse to $\bm v_n-\bm v_ L$ is the Kopnin--Kravtsov force. Uniting this force with the Magnus force in the left-hand side of \eq{vd} at $T=0$ ($n_s=n$) one obtains the total transverse force
\bem
\bm F_\perp=m n  \kappa_c[\hat z \times (\bm v_L-\bm v_s)]-m n \kappa_c\frac{[\hat z \times(\bm v_L-\bm v_ n)] }{1+\omega_0^2\tau^2}
\nonumber \\
=m n_M   \kappa_c[\hat z \times \bm v_L]-m n \kappa_c \left\{ [\hat z \times \bm v_s]- \frac{[\bm v_ n\times \hat z] }{1+\omega_0^2\tau^2}\right\},
     \eem{}
where
\be
n_M = \frac{\omega_0^2\tau^2 }{1+\omega_0^2\tau^2}
     \ee{nM}
is the density determining the effective Magnus force on the vortex.     
In the limit $\omega_0\tau \to 0$  the Kopnin--Kravtsov force compensates the Magnus force, and the total transverse force vanishes.  

For better understanding of the Kopnin--Kravtsov force let us derive it by replacing in the integral of \eq{colInt} 
the collision term by the left-hand side of the Boltzmann equation (\ref{kinO}),  which is the divergence of the flow of quasiparticles in the phase space $\{\alpha,~L_z\}$ and corresponds to 
 the  Liouville equation. After integration  by parts the Kopnin--Kravtsov force is 
\bem
\bm F_c = {1\over 2}\int \limits_{-k_F}^{k_F}{dk_z\over 2\pi} \int d\alpha
 \int\limits_{L_z^{min}} ^{L_z^{max}}{dL_z \over 2\pi}\bm k
 \nonumber \\\times
  \left(- {\partial\epsilon \over \partial \alpha }{\partial f\over \partial L_z}+{\partial\epsilon \over \partial L_z}{\partial f\over \partial \alpha} \right)
 = -{1\over 2}\int \limits_{-k_F}^{k_F}{dk_z\over 2\pi} \int d\alpha
 \nonumber \\
\times  \left\{ \left.\bm k {\partial\epsilon \over \partial \alpha }f( L_z)\right |_{L_z^{min}}^{L_z^{max}}+ \int\limits_{L_z^{min}} ^{L_z^{max}}{dL_z \over 2\pi}[\hat z\times \bm k]\omega_0 f(L_z) \right\}. \nonumber \\
  \eem{}   
Here we took into account that $\partial \epsilon/\partial L_z$ does not depend on $\alpha$ and $\partial \epsilon/\partial \alpha$ does not depend on $L_z$. The first term in the final expression is the momentum flux 
in the $L_z$ space 
through the upper and the lower boundaries of the gap and the second term is the momentum transfer from the external force driving the vortex at the process of the bound state rotation with the angular velocity $\omega_0=\partial\epsilon /\partial L_z$. While the isotropic part of the distribution function contributes to the first term, only the anisotropic part provides the second term. Restricting ourselves with the case of $\omega_0\tau\to 0$ when the solution of the Boltzmann equation $f=f_n$ is given by Eq.~ (\ref{fn})   one obtains the Kopnin--Kravtsov force for this case. So the origin of the Kopnin--Kravtsov force looks clear and does not require a reference to the artificial concept of spectral flow. 

\section{Effect of vortex mass on vortex dynamics} \label{VorIn}

Taking into account all forces discussed above the general equation describing free motion of the vortex in the resting liquid ($\bm v_s=\bm v_n=0$) is 
\be
\mu_v {d \bm v_L\over dt} -m n_M \kappa \left[\hat z \times   \bm v_L\right] =- \gamma \bm v_L - \mu_\perp \left[\hat z \times{d \bm v_L\over dt}\right], 
  \ee{vm}
where  $n_M$ is given by \eq{nM} and $\kappa$ must be replaced by $\kappa_c$ in the Fermi superfluid. At zero temperature  $n_M$ varies from $n_M=n$ for superclean superconductors down to $n_M=0$ for moderately dirty superconductors. The right-hand side of the equation contains two dissipative forces. The second of them is connected with the transverse vortex mass originated from core bound states. \cite{KopV}  It determines a high-frequency  correction to the dissipative (longitudinal) mutual-friction force, which does not appear in the Hamiltonian but has its counterpart in the dissipative function. In order to see it let us  derive the time variation of the kinetic energy of the vortex:
\be 
{dE\over dt} ={d\over dt}\left(\mu_v {  v_L^2 \over 2}\right)=- 2F_D.
   \ee{}
Here the dissipative function is
\be 
F_D= {\gamma v_L^2\over 2}+{\mu_\perp\over 2}  \left[ \bm v_L\times  {d\bm v_L\over dt}\right]\cdot \hat z.
    \ee{}
The contribution of the transverse mass to the dissipative function is not positively defined. Therefore, the equation of motion as given by \eq{vm} makes sense only if the transverse-mass contribution is small compared to that of the usual friction force $\propto\gamma$.

Without dissipation  \eq{vm} is analogous to the equation of motion of a charged particle in a magnetic field. The vortex rotates around a circular orbit with the angular velocity $\omega_c =mn_M \kappa /\mu_v $, which is an analog of the cyclotron frequency. The frequency $\omega_c$ also characterizes the frequency of an a.c. process at which the vortex-mass  effect can compete with the transverse Magnus force. In the Bose liquid with $n_M=n$ and  the vortex mass $\mu_v \sim mn r_c^2 $  the frequency $\omega_c$ is on the order of $c_s^2/\kappa \sim \kappa/r_c^2$. A phonon with such a high frequency has a wavelength comparable with the core radius $r_c$. If the vortex moves around a  circumference of the radius $r_0$, which exceeds the core radius, the linear velocity $\omega_c r_0$  exceeds the value of the critical velocity $c_s \sim \kappa / r_c$. Hardly this rotation is of practical importance.  In the Fermi liquid the frequency $\omega_c$ is of the same order as the frequency $\omega_0$, which determines the distance $\hbar\omega_0$ between core energy levels. Though the latter is small in comparison with the gap $\Delta$ in the weak-coupling limit, the frequency itself is rather high. This is true both in the pure limit when $n_M=n$ and  $\mu_v \sim mn r_c^2 $ and in the dirty limit when  $n_M=\omega_0^2 \tau^2 n$ and  $\mu_v \sim \omega_0^2\tau^2mn r_c^2 $. In all, it is not simple to reveal the vortex mass in superfluids and  superconductors, though some experimental evidence of the vortex mass in superconducting thin films has been recently reported.\cite{Pol}

\section{Discussion and conclusions} \label{spFl}

Since our analysis does not reveal the spectral flow in the core of the moving vortex let us discuss the arguments by Volovik \cite{Vol93,VolB} in favor of its existence.
Deriving the spectral flow Volovik considered  the angular momentum  $L'_z= \hat z \cdot[ (\bm r- (\bm v_L-\bm v_n) t) \times \bm p]$ around the axis, which moves together with the thermal bath (normal component). 
Here $\bm r$ is the position vector with the origin on the symmetry axis of the moving vortex.
Volovik's  angular momentum varies in time:
\be
{dL'_z\over dt}=-\hat z \cdot[  (\bm v_L-\bm v_n)  \times \bm p].
   \ee{V}        
Since the energy of the bound state is proportional to the angular momentum, Volovik concluded that the energy levels move in the energy space and cross the zero energy level with the rate proportional to $\bm v_L-\bm v_n$. The problem with this argument is that the position of the bound state energy with respect to the gap depends on the angular momentum about the symmetry axis of the vortex in the coordinate frame moving together with the vortex. Then the angular momentum is conserved and provides a good  quantum number, which determines the energy of the bound state. In any other coordinate frame with the reference axis, which does not coincides with the vortex axis, the angular momentum is not conserved and is not a quantum number. Moreover, deriving \eq{V}, Volovik assumed that the momentum $\bm p$ of the bound state does not varies in time. Meanwhile, in a genuine bound state the momentum $\bm p$ rotates with the angular velocity $\omega_0$ and vanishes in average. As a result, d$L'_z/ dt$ vanishes also, and the angular momentum determined with reference to any axis does not  differ in average from the angular momentum around the vortex symmetry axis. This is a direct consequence of the theorem of mechanics, which tells that for a system with vanishing velocity of the center of mass the angular momentum does not depend on the choice of the reference axis.  So the vortex motion with respect to the thermal bath does not lead to the spectral flow.

\begin{figure}
\includegraphics[width=.4\textwidth]{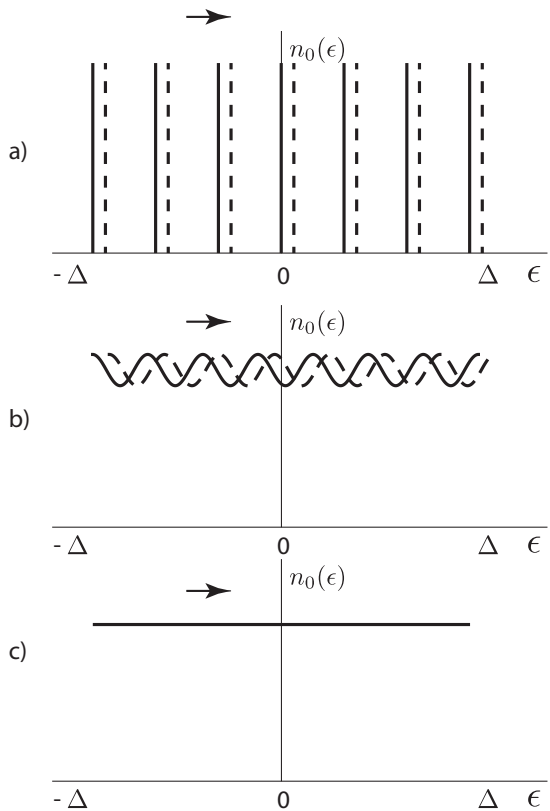}
\caption[]{Effect of shift of energy levels on the density of states $n_0(\epsilon)$ at various $\omega_0\tau$.  The density of states is shown by solid lines before the shift and by dashed lines after the shift. a) $\omega_0\tau \to \infty$. The density of states is a chain of sharped peaks. b) $\omega_0\tau \ll 1$. Very broad peaks strongly overlap and cause only weak oscillations of  the density of states, which are still noticeable  in principle. c) $\omega_0\tau=0$. The plot of the density of states is totally flat and its shift does not lead to any physical consequence.}
\label{DOS}
\end{figure}

Volovik stressed that his derivation was only for continuum limit $\omega_0 \tau \to 0$ when levels are strongly broadened and in fact cease to be discrete levels. Originally the spectral flow concept was considered  only for discrete levels. 
In the continuum limit the very concept of the spectral flow becomes ambiguous. This is illustrated in Fig.~\ref{DOS}, which shows the effect of the level shift on the density of states $n_0(\epsilon)$ for various $\omega_0\tau$. Without collisions  ($\omega_0 \tau \to \infty$) the density of states is a chain of very narrow peaks and a shift of the levels with respect to the forbidden gap is a clear effect (Fig.~\ref{DOS}a). For very small but still finite $\omega_0\tau$ the effect of level shift  on the density of states is much weaker but still noticeable (Fig.~\ref{DOS}b). In  the extreme case $\omega_0\tau=0$ when oscillations of the density of states are totally undetectable the level shift does not lead to any effect and cannot influence any physical process. Without taking into account whatever tiny oscillations of the density of states it is impossible even to define it.  

Altogether this  puts in question not the Kopnin--Kravtsov force itself but the connection of the force with the spectral flow. So the claim that the experiment on mutual friction force confirms the spectral flow\cite{BevN,VolB} is not justified. It is  the Kopnin--Kravtsov force, which was revealed in the experiment, but not the spectral flow.

Arguing for the spectral flow in vortex dynamics they frequently draw an analogy with the Andreev-reflection bound states in the Superconductor--Normal metal--Superconductor (SNS) junctions. Though this analogy is useful indeed for the bound states in the model of normal core \cite{stone96} it fails with respect to the role of spectral flow. In the SNS system the spectral flow really exists if the phase difference between two superconducting banks monotonously varies in time, as,  e.g., in   the a.c. Josephson effect. But the superfluid phase difference across the core of the moving vortex   does not vary in time in average. Therefore, the spectral flow exists in the former case, but is totally absent in the latter.

The absence of spectral flow in the core automatically rules out the spectral flow in the continuum of delocalized states also suggested by \citet{VolB}. Indeed, in stationary processes these spectral flows should be equal. Otherwise there were accumulation or depletion of states at the borders between localized and delocalized states. It is shown in this paper that the whole transverse force on the vortex from delocalized states in Fermi superfluids can be explained by peculiarities of the Aharonov--Bohm effect for BCS quasiparticles without referring to the spectral flow.

The paper clarifies also the role of the backflow on the vortex mass. The backflow is an ubiquitous phenomenon, which arises from mismatching of currents inside and outside the vortex core, either due to suppression of the fluid density in the Bose liquid, or due to to currents through core bound states in the Fermi liquid. Its existence follows from the conservation law for the particle number (charge). In the Fermi liquid the backflow leads to renormalization of the Kopnin vortex mass by a numerical factor  both for singular and continuous vortices.

\begin{acknowledgments}
I thank Vadim Gurevich, Nikolai Kopnin, Andrei Shelankov,  and Michael Stone for interesting discussions.
The work was supported by the grant of the Israel Academy of Sciences and Humanities and by the FP7 program Microkelvin of the European Union.
\end{acknowledgments}

\appendix
\section{Bound states in a core with linear growth of superfluid density} \label{bss}

We consider a quasiparticle inside the core, which moves back and forth along an approximately straight trajectory parallel to the $y$ changing its direction of motion via Andreev reflection. One can refer to the Bogolyubov--de Gennes equations  in the 1D case eliminating fast oscillations of the wave function by the transformation $u=\tilde u e^{ik_F y}$,  $v=\tilde v e^{ik_F y}$. For the sake of  simplicity we assume that there no component of the wave vector parallel to the $z$ axis.
Neglecting second derivatives of $\tilde u$ and $\tilde v$ the Bogolyubov--de Gennes equations are
\bem
-i\hbar v_F  {d \tilde  u (b, y)\over dy} + 
{\Delta  r\over r_c} e^{i\theta} \tilde   v(b, y) =\epsilon \tilde  u(b, y),
\nonumber \\
i\hbar v_F  {d \tilde  v (b, y)\over dy} + 
{\Delta  r\over r_c} e^{-i\theta} \tilde   u(b, y) =\epsilon \tilde  v(b, y).
     \eem{Cv}
Here $r=\sqrt{b^2+y^2}$, and the linear dependence of the gap $\Delta(r) =\Delta r/r_c$ on the distance $r$ is assumed.  In  the absence of superfluid motion through the core
the phase $\theta$ coincides with the azimuthal angle $\phi= \arctan(y/b)$, and the Bogolyubov--de Gennes equations become
\bem
-i\hbar v_F  {d \tilde  u (b, y)\over dy} + 
{\Delta  (b+iy)\over r_c} \tilde   v(b, y) =\epsilon \tilde  u(b, y),
\nonumber \\
i\hbar v_F  {d \tilde  v (b, y)\over dy} + 
{\Delta  (b-iy)\over r_c}  \tilde   u(b, y) =\epsilon \tilde  v(b, y).
     \eem{Cuv}
The normalized solution of the Bogolyubov--de Gennes equations is 
\be
\tilde u=-\tilde v={e^{- y^2/2 \tilde r^2}\over 2\sqrt{\pi}\tilde r}
    \ee{wfC} 
with the energy equal to 
\be
     \epsilon_0= - {b  \Delta \over r_c}= - \omega_0 L_z.
                 \ee{en1}
Here  the length $\tilde r =\sqrt{r_c\xi_c}$ is the geometric average of the core radius $r_c $ and  the coherence length $\xi_c=\hbar v_F/\Delta $ with all three lengths being of the same order of magnitude, $L_z=\hbar k_F b$ is  the angular momentum of the bound state,  and the frequency
 \be
\omega_0=  {  \Delta \over \hbar k_F r_c}
                 \ee{}
gives the angular velocity of slow trajectory rotation around the vortex axis, in accordance with the canonical relation equating the rotation velocity to $\partial {\cal H}/\partial L_z$.

The energy spectrum given by \eq{en1} insignificantly differs from the spectrum obtained in the original paper \cite{corSt} and in the book by \citet{deGen} more accurately using the partial-wave analysis and a more realistic variation of the gap $\Delta$ in the space. This agreement confirms a simple picture of the bound states assuming well defined trajectories of quasiparticle motion. However, it is necessary to stress that though the trajectory is well defined in the sense that the impact parameter is well defined, the motion along trajectory cannot be described semiclassically. In particular, our solution shows that there are no well defined Andreev-reflection points. Using the semiclassical approach and  the Bohr--Sommerfeld condition for calculation of energy levels one obtains a totally wrong spectrum, which is not linear in the angular momentum. So the semiclassical theory of motion along the trajectory of the bound state is valid only for the model of the totally normal core but not for more realistic models with non-zero order parameter in the core.

\section{Vortex mass of a core with linear growth of superfluid density in the Bose superfluid} \label{bm}

 According to the Gross--Pitaevskii theory in the vortex core  the density grows linearly with the distance $r$ from the vortex axis. Extrapolating this dependence up to the core radius $r_c$ and approximating the density outside the core by the constant value $n$, the continuous density distribution is
\be
n(r)=\left\{ 
\begin{array}{cc}
n{r\over r_c}  &  \mbox{at}~r<r_c    \\
n  &       \mbox{at}~r>r_c
\end{array} \right.  .
   \ee{} 
The liquid mass inside the core,  
\be
\tilde m_{core}=2\pi{mn\over r_c}\int\limits_0^{r_c} r^2\, dr={2\over 3}\mu_{core}, 
  \ee{}
is by the factor 2/3 less than the core mass $\mu_{core}$ estimated under the assumption that the liquid density is not suppressed inside the core.
If the superfluid moves past  the vortex the continuity equation in the coordinate frame related to the vortex is 
\be
\bm \nabla [n(r) \bm \nabla \theta]= n(r)  \bm \nabla^2 \theta +\bm \nabla n(r) \cdot  \bm \nabla \theta=0,
     \ee{}
where the phase $\theta$ determines the velocity field: $\bm v_s(\bm r)- {\bm v}_L=(\kappa/2\pi)\bm \nabla \theta(\bm r)$. From symmetry all fields are dipole fields, and the phase in the cylindric coordinate system is $ \theta(\bm r)= \theta( r) \cos \phi$, where $\phi$ is the azimuthal angle with respect to the velocity $\bm v_s-\bm v_L$. The one-dimensional function $\theta(r)$ is determined from the equation:
\be
{d\theta ^2 \over dr^2} + \left[{1\over r}+{1\over n(r)}{dn(r)  \over dr}\right]{d\theta  \over dr}-{\theta\over r^2}=0.
     \ee{gp-c}
Inside the core \eq{gp-c} yields that $ \theta \propto r^\alpha$ with the exponent $\alpha =(\sqrt{5}-1)/2<1$. This means that the velocity (but not the current!) has a weak integrable singularity at $r=0$. The continuity of the azimuthal component of the superfluid velocity at the core boundary $r=r_c$ is satisfied by the following phase distribution outside and inside the core (the azimuthal angle dependence is omitted):
\be 
\theta_{out}={2\pi\over \kappa}\left(v_s r -{v_{bf}r_c^2\over r}\right),~~ \theta_{in}={2\pi\over \kappa}(v_s -v_{bf} ) {r^\alpha\over r_c^{\alpha-1}}.
              \ee{}
Here $v_s$ is the superfluid velocity far from the vortex in the coordinate frame moving with the vortex and $v_{bf}$ is the amplitude of the backflow velocity field given by  \eq{bf}. Continuity of the radial velocity gives the condition:
\be
{d\theta_{in}\over dr}={2\pi\over \kappa}\alpha (v_s -v_{bf} )={d\theta_{out}\over dr}={2\pi\over \kappa}(v_{bf}+v_s).
           \ee{}
 This yields the relation
\be
v_{bf}=-v_s\frac{1-\alpha}  {1+\alpha} .
    \ee{}
The total momentum  includes the momentum $\bm P_{in}$ inside the core, the momentum of transport superfluid velocity $\bm v_s$ outside the core,
 and the Kelvin impulse of the backflow velocity field outside the core [\eq{Kel}]:
\bem
\bm P_L={m\kappa\over 2\pi} \int \limits_{r<r_c} n(r)\bm \nabla \theta_{in}\,d\bm r+ mn(S-\pi r_c^2) (\bm v_s -\bm v_L) 
\nonumber \\ + \bm P_K=mn \left [S- \pi r_c^2\left({\alpha \over 2+\alpha}+\frac{1-\alpha}  {1+\alpha}\right)\right] (\bm v_s -\bm v_L),  
\nonumber \\ 
     \eem{}
where the superfluid velocity $\bm v_s$ was replaced by the relative velocity $\bm v_s -\bm v_L$. The last step is to transform the momentum $\bm P_L$ in the coordinate frame moving with the vortex to the momentum in the arbitrary coordinate frame:
\bem
\bm P=\bm P_L+ [mn(S-\pi r_c^2)+\tilde m_{core}] \bm v_L
\nonumber \\ 
=mn \left [S- \pi r_c^2\left({\alpha \over 2+\alpha}+\frac{1-\alpha}  {1+\alpha}\right)\right]  {\bm v}_s +\mu_v \bm v_L.
     \eem{}
Here $\mu_v$ is the vortex mass. Taking into account the value $\alpha =(\sqrt{5}-1)/2$ the vortex mass is  
\bem
\mu_v= \pi r_c^2mn \left[  {\alpha \over 2+\alpha} +\frac{1-\alpha}  {1+\alpha}-{1\over 3}\right]
\nonumber \\
=\pi r_c^2mn\left[2\sqrt{5} -4{1\over 3}\right]=0.139\mu_{core}.
    \eem{}


\begin{thebibliography}{39}%
\makeatletter
\providecommand \@ifxundefined [1]{%
 \@ifx{#1\undefined}
}%
\providecommand \@ifnum [1]{%
 \ifnum #1\expandafter \@firstoftwo
 \else \expandafter \@secondoftwo
 \fi
}%
\providecommand \@ifx [1]{%
 \ifx #1\expandafter \@firstoftwo
 \else \expandafter \@secondoftwo
 \fi
}%
\providecommand \natexlab [1]{#1}%
\providecommand \enquote  [1]{``#1''}%
\providecommand \bibnamefont  [1]{#1}%
\providecommand \bibfnamefont [1]{#1}%
\providecommand \citenamefont [1]{#1}%
\providecommand \href@noop [0]{\@secondoftwo}%
\providecommand \href [0]{\begingroup \@sanitize@url \@href}%
\providecommand \@href[1]{\@@startlink{#1}\@@href}%
\providecommand \@@href[1]{\endgroup#1\@@endlink}%
\providecommand \@sanitize@url [0]{\catcode `\\12\catcode `\$12\catcode
  `\&12\catcode `\#12\catcode `\^12\catcode `\_12\catcode `\%12\relax}%
\providecommand \@@startlink[1]{}%
\providecommand \@@endlink[0]{}%
\providecommand \url  [0]{\begingroup\@sanitize@url \@url }%
\providecommand \@url [1]{\endgroup\@href {#1}{\urlprefix }}%
\providecommand \urlprefix  [0]{URL }%
\providecommand \Eprint [0]{\href }%
\providecommand \doibase [0]{http://dx.doi.org/}%
\providecommand \selectlanguage [0]{\@gobble}%
\providecommand \bibinfo  [0]{\@secondoftwo}%
\providecommand \bibfield  [0]{\@secondoftwo}%
\providecommand \translation [1]{[#1]}%
\providecommand \BibitemOpen [0]{}%
\providecommand \bibitemStop [0]{}%
\providecommand \bibitemNoStop [0]{.\EOS\space}%
\providecommand \EOS [0]{\spacefactor3000\relax}%
\providecommand \BibitemShut  [1]{\csname bibitem#1\endcsname}%
\let\auto@bib@innerbib\@empty
\bibitem [{\citenamefont {Sonin}(1987)}]{RMP}%
  \BibitemOpen
  \bibfield  {author} {\bibinfo {author} {\bibfnamefont {E.~B.}\ \bibnamefont
  {Sonin}},\ }\href@noop {} {\bibfield  {journal} {\bibinfo  {journal} {Rev.
  Mod. Phys.}\ }\textbf {\bibinfo {volume} {59}},\ \bibinfo {pages} {87}
  (\bibinfo {year} {1987})}\BibitemShut {NoStop}%
\bibitem [{\citenamefont {Kopnin}(2001)}]{Kop}%
  \BibitemOpen
  \bibfield  {author} {\bibinfo {author} {\bibfnamefont {N.~B.}\ \bibnamefont
  {Kopnin}},\ }\href@noop {} {\emph {\bibinfo {title} {Theory of Nonequilibrium
  Superconductivity}}}\ (\bibinfo  {publisher} {Oxford University Press},\
  \bibinfo {year} {2001})\BibitemShut {NoStop}%
\bibitem [{\citenamefont {Kopnin}(2002)}]{KopR}%
  \BibitemOpen
  \bibfield  {author} {\bibinfo {author} {\bibfnamefont {N.~B.}\ \bibnamefont
  {Kopnin}},\ }\href@noop {} {\bibfield  {journal} {\bibinfo  {journal} {Rep.
  Prog. Phys.}\ }\textbf {\bibinfo {volume} {65}},\ \bibinfo {pages} {1633}
  (\bibinfo {year} {2002})}\BibitemShut {NoStop}%
\bibitem [{\citenamefont {Sonin}(2002)}]{Magn}%
  \BibitemOpen
  \bibfield  {author} {\bibinfo {author} {\bibfnamefont {E.~B.}\ \bibnamefont
  {Sonin}},\ }in\ \href@noop {} {\emph {\bibinfo {booktitle} {Vortices in
  Unconventional Superconductors and Superfluids}}},\ \bibinfo {editor} {edited
  by\ \bibinfo {editor} {\bibfnamefont {R.~P.}\ \bibnamefont {Huebener}},
  \bibinfo {editor} {\bibfnamefont {N.}~\bibnamefont {Schopohl}}, \ and\
  \bibinfo {editor} {\bibfnamefont {G.~E.}\ \bibnamefont {Volovik}}}\ (\bibinfo
   {publisher} {Springer-Verlag},\ \bibinfo {year} {2002})\ pp.\ \bibinfo
  {pages} {119--145}\BibitemShut {NoStop}%
\bibitem [{\citenamefont {Shelankov}(2002)}]{ShelB}%
  \BibitemOpen
  \bibfield  {author} {\bibinfo {author} {\bibfnamefont {A.~L.}\ \bibnamefont
  {Shelankov}},\ }in\ \href@noop {} {\emph {\bibinfo {booktitle} {Vortices in
  Unconventional Superconductors and Superfluids}}},\ \bibinfo {editor} {edited
  by\ \bibinfo {editor} {\bibfnamefont {R.~P.}\ \bibnamefont {Huebener}},
  \bibinfo {editor} {\bibfnamefont {N.}~\bibnamefont {Schopohl}}, \ and\
  \bibinfo {editor} {\bibfnamefont {G.~E.}\ \bibnamefont {Volovik}}}\ (\bibinfo
   {publisher} {Springer-Verlag},\ \bibinfo {year} {2002})\ pp.\ \bibinfo
  {pages} {147---166}\BibitemShut {NoStop}%
\bibitem [{\citenamefont {Volovik}(2003)}]{VolB}%
  \BibitemOpen
  \bibfield  {author} {\bibinfo {author} {\bibfnamefont {G.~E.}\ \bibnamefont
  {Volovik}},\ }\href@noop {} {\emph {\bibinfo {title} {The Universe in a
  Helium Droplet}}}\ (\bibinfo  {publisher} {Oxford University Press},\
  \bibinfo {year} {2003})\BibitemShut {NoStop}%
\bibitem [{\citenamefont {Suhl}(1965)}]{Suhl}%
  \BibitemOpen
  \bibfield  {author} {\bibinfo {author} {\bibfnamefont {H.}~\bibnamefont
  {Suhl}},\ }\href@noop {} {\bibfield  {journal} {\bibinfo  {journal} {Phys.
  Rev. Lett.}\ }\textbf {\bibinfo {volume} {14}},\ \bibinfo {pages} {226}
  (\bibinfo {year} {1965})}\BibitemShut {NoStop}%
\bibitem [{\citenamefont {Popov}(1973)}]{VNP}%
  \BibitemOpen
  \bibfield  {author} {\bibinfo {author} {\bibfnamefont {V.~N.}\ \bibnamefont
  {Popov}},\ }\href@noop {} {\bibfield  {journal} {\bibinfo  {journal} {Zh.
  Eksp. Teor. Fiz}\ }\textbf {\bibinfo {volume} {64}},\ \bibinfo {pages} {672}
  (\bibinfo {year} {1973})},\ \bibinfo {note} {[Sov. Phys.--JETP {\bf 37}, 341
  (1973)]}\BibitemShut {NoStop}%
\bibitem [{\citenamefont {Kopnin}(1978)}]{KopM}%
  \BibitemOpen
  \bibfield  {author} {\bibinfo {author} {\bibfnamefont {N.~B.}\ \bibnamefont
  {Kopnin}},\ }\href@noop {} {\bibfield  {journal} {\bibinfo  {journal} {Pis'ma
  Zh. Eksp. Teor. Fiz.}\ }\textbf {\bibinfo {volume} {27}},\ \bibinfo {pages}
  {417} (\bibinfo {year} {1978})},\ \bibinfo {note} {[JETP Lett., {\bf 23}, 578
  (1976)]}\BibitemShut {NoStop}%
\bibitem [{\citenamefont {Baym}\ and\ \citenamefont {Chandler}(1983)}]{Bay83}%
  \BibitemOpen
  \bibfield  {author} {\bibinfo {author} {\bibfnamefont {G.}~\bibnamefont
  {Baym}}\ and\ \bibinfo {author} {\bibfnamefont {E.}~\bibnamefont
  {Chandler}},\ }\href@noop {} {\bibfield  {journal} {\bibinfo  {journal} {J.
  Low Temp. Phys.}\ }\textbf {\bibinfo {volume} {50}},\ \bibinfo {pages} {57}
  (\bibinfo {year} {1983})}\BibitemShut {NoStop}%
\bibitem [{\citenamefont {Duan}\ and\ \citenamefont {Leggett}(1992)}]{DL}%
  \BibitemOpen
  \bibfield  {author} {\bibinfo {author} {\bibfnamefont {J.-M.}\ \bibnamefont
  {Duan}}\ and\ \bibinfo {author} {\bibfnamefont {A.~J.}\ \bibnamefont
  {Leggett}},\ }\href@noop {} {\bibfield  {journal} {\bibinfo  {journal} {Phys.
  Rev. Lett.}\ }\textbf {\bibinfo {volume} {68}},\ \bibinfo {pages} {1216}
  (\bibinfo {year} {1992})},\ \bibinfo {note} {er., {\sl ibid} {\bf 69}, 1148
  (1992)}\BibitemShut {NoStop}%
\bibitem [{\citenamefont {Duan}(1993)}]{Duan93}%
  \BibitemOpen
  \bibfield  {author} {\bibinfo {author} {\bibfnamefont {J.-M.}\ \bibnamefont
  {Duan}},\ }\href@noop {} {\bibfield  {journal} {\bibinfo  {journal} {Phys.
  Rev. B}\ }\textbf {\bibinfo {volume} {48}},\ \bibinfo {pages} {333} (\bibinfo
  {year} {1993})}\BibitemShut {NoStop}%
\bibitem [{\citenamefont {Duan}(1994)}]{Duan}%
  \BibitemOpen
  \bibfield  {author} {\bibinfo {author} {\bibfnamefont {J.-M.}\ \bibnamefont
  {Duan}},\ }\href@noop {} {\bibfield  {journal} {\bibinfo  {journal} {Phys.
  Rev. B}\ }\textbf {\bibinfo {volume} {49}},\ \bibinfo {pages} {12381}
  (\bibinfo {year} {1994})}\BibitemShut {NoStop}%
\bibitem [{\citenamefont {Sonin}\ \emph {et~al.}(1998)\citenamefont {Sonin},
  \citenamefont {Geshkenbein}, \citenamefont {{van Otterlo}},\ and\
  \citenamefont {Blatter}}]{Son98}%
  \BibitemOpen
  \bibfield  {author} {\bibinfo {author} {\bibfnamefont {E.~B.}\ \bibnamefont
  {Sonin}}, \bibinfo {author} {\bibfnamefont {V.~B.}\ \bibnamefont
  {Geshkenbein}}, \bibinfo {author} {\bibfnamefont {A.}~\bibnamefont {{van
  Otterlo}}}, \ and\ \bibinfo {author} {\bibfnamefont {G.}~\bibnamefont
  {Blatter}},\ }\href@noop {} {\bibfield  {journal} {\bibinfo  {journal} {Phys.
  Rev B}\ }\textbf {\bibinfo {volume} {57}},\ \bibinfo {pages} {575} (\bibinfo
  {year} {1998})}\BibitemShut {NoStop}%
\bibitem [{\citenamefont {Kopnin}\ and\ \citenamefont {Vinokur}(1998)}]{KopV}%
  \BibitemOpen
  \bibfield  {author} {\bibinfo {author} {\bibfnamefont {N.~B.}\ \bibnamefont
  {Kopnin}}\ and\ \bibinfo {author} {\bibfnamefont {V.~M.}\ \bibnamefont
  {Vinokur}},\ }\href@noop {} {\bibfield  {journal} {\bibinfo  {journal} {Phys.
  Rev. Lett.}\ }\textbf {\bibinfo {volume} {81}},\ \bibinfo {pages} {3952}
  (\bibinfo {year} {1998})}\BibitemShut {NoStop}%
\bibitem [{\citenamefont {Volovik}(1993)}]{Vol93}%
  \BibitemOpen
  \bibfield  {author} {\bibinfo {author} {\bibfnamefont {G.~E.}\ \bibnamefont
  {Volovik}},\ }\href@noop {} {\bibfield  {journal} {\bibinfo  {journal}
  {Pis'ma Zh. Eksp. Teor. Fiz.}\ }\textbf {\bibinfo {volume} {57}},\ \bibinfo
  {pages} {233} (\bibinfo {year} {1993})},\ \bibinfo {note} {[JETP Letters,
  {\bf 57}, 244 (1993)]}\BibitemShut {NoStop}%
\bibitem [{\citenamefont {Bevan}\ \emph {et~al.}(1997)\citenamefont {Bevan},
  \citenamefont {Manninen}, \citenamefont {Cook}, \citenamefont {Hook},
  \citenamefont {Hall}, \citenamefont {Vachaspati},\ and\ \citenamefont
  {Volovik}}]{BevN}%
  \BibitemOpen
  \bibfield  {author} {\bibinfo {author} {\bibfnamefont {T.~D.~C.}\
  \bibnamefont {Bevan}}, \bibinfo {author} {\bibfnamefont {A.~J.}\ \bibnamefont
  {Manninen}}, \bibinfo {author} {\bibfnamefont {J.~B.}\ \bibnamefont {Cook}},
  \bibinfo {author} {\bibfnamefont {J.~R.}\ \bibnamefont {Hook}}, \bibinfo
  {author} {\bibfnamefont {H.~E.}\ \bibnamefont {Hall}}, \bibinfo {author}
  {\bibfnamefont {T.}~\bibnamefont {Vachaspati}}, \ and\ \bibinfo {author}
  {\bibfnamefont {G.~E.}\ \bibnamefont {Volovik}},\ }\href@noop {} {\bibfield
  {journal} {\bibinfo  {journal} {Nature}\ }\textbf {\bibinfo {volume} {386}},\
  \bibinfo {pages} {689} (\bibinfo {year} {1997})}\BibitemShut {NoStop}%
\bibitem [{\citenamefont {Booss-Bavnbek}\ and\ \citenamefont
  {Wojciechowski}(1993)}]{spFl}%
  \BibitemOpen
  \bibfield  {author} {\bibinfo {author} {\bibfnamefont {B.}~\bibnamefont
  {Booss-Bavnbek}}\ and\ \bibinfo {author} {\bibfnamefont {K.~P.}\ \bibnamefont
  {Wojciechowski}},\ }\href@noop {} {\emph {\bibinfo {title} {Elliptic Boundary
  Problems for Dirac Operators}}}\ (\bibinfo  {publisher} {Birkh{\"a}user},\
  \bibinfo {year} {1993})\ \bibinfo {note} {ch. 17}\BibitemShut {NoStop}%
\bibitem [{\citenamefont {Stone}(1996)}]{stone96}%
  \BibitemOpen
  \bibfield  {author} {\bibinfo {author} {\bibfnamefont {M.}~\bibnamefont
  {Stone}},\ }\href@noop {} {\bibfield  {journal} {\bibinfo  {journal} {Phys.
  Rev. B}\ }\textbf {\bibinfo {volume} {54}},\ \bibinfo {pages} {13222}
  (\bibinfo {year} {1996})}\BibitemShut {NoStop}%
\bibitem [{\citenamefont {Caroli}\ \emph {et~al.}(1964)\citenamefont {Caroli},
  \citenamefont {de~Gennes},\ and\ \citenamefont {Matricon}}]{corSt}%
  \BibitemOpen
  \bibfield  {author} {\bibinfo {author} {\bibfnamefont {C.}~\bibnamefont
  {Caroli}}, \bibinfo {author} {\bibfnamefont {P.~G.}\ \bibnamefont
  {de~Gennes}}, \ and\ \bibinfo {author} {\bibfnamefont {J.}~\bibnamefont
  {Matricon}},\ }\href@noop {} {\bibfield  {journal} {\bibinfo  {journal}
  {Physics Letters}\ }\textbf {\bibinfo {volume} {9}},\ \bibinfo {pages} {307}
  (\bibinfo {year} {1964})}\BibitemShut {NoStop}%
\bibitem [{\citenamefont {de~Gennes}(1966)}]{deGen}%
  \BibitemOpen
  \bibfield  {author} {\bibinfo {author} {\bibfnamefont {P.~G.}\ \bibnamefont
  {de~Gennes}},\ }\href@noop {} {\emph {\bibinfo {title} {Superconductivity of
  metals and alloys}}}\ (\bibinfo  {publisher} {Benjamin},\ \bibinfo {year}
  {1966})\BibitemShut {NoStop}%
\bibitem [{\citenamefont {Kopnin}\ and\ \citenamefont
  {Kravtsov}(1976{\natexlab{a}})}]{Kop76b}%
  \BibitemOpen
  \bibfield  {author} {\bibinfo {author} {\bibfnamefont {N.~B.}\ \bibnamefont
  {Kopnin}}\ and\ \bibinfo {author} {\bibfnamefont {V.~E.}\ \bibnamefont
  {Kravtsov}},\ }\href@noop {} {\bibfield  {journal} {\bibinfo  {journal}
  {Pis'ma Zh. Eksp. Teor. Fiz.}\ }\textbf {\bibinfo {volume} {23}},\ \bibinfo
  {pages} {631} (\bibinfo {year} {1976}{\natexlab{a}})},\ \bibinfo {note}
  {[{JETP Lett.}, {\bf 23}, 578 (1976)]}\BibitemShut {NoStop}%
\bibitem [{\citenamefont {Lifshitz}\ and\ \citenamefont
  {Pitaevskii}(1957)}]{Lif57}%
  \BibitemOpen
  \bibfield  {author} {\bibinfo {author} {\bibfnamefont {E.~M.}\ \bibnamefont
  {Lifshitz}}\ and\ \bibinfo {author} {\bibfnamefont {L.~P.}\ \bibnamefont
  {Pitaevskii}},\ }\href@noop {} {\bibfield  {journal} {\bibinfo  {journal}
  {Zh. Eksp. Teor. Fiz.}\ }\textbf {\bibinfo {volume} {33}},\ \bibinfo {pages}
  {535} (\bibinfo {year} {1957})},\ \bibinfo {note} {[{Sov. Phys.--JETP},
  {\bf 6}, 418 (1957)]}\BibitemShut {NoStop}%
\bibitem [{\citenamefont {Sonin}(1975)}]{Son75}%
  \BibitemOpen
  \bibfield  {author} {\bibinfo {author} {\bibfnamefont {E.~B.}\ \bibnamefont
  {Sonin}},\ }\href@noop {} {\bibfield  {journal} {\bibinfo  {journal} {Zh.
  Eksp. Teor. Fiz.}\ }\textbf {\bibinfo {volume} {69}},\ \bibinfo {pages} {921}
  (\bibinfo {year} {1975})},\ \bibinfo {note} {[{Sov. Phys.--JETP}, {\bf
  42}, 469 (1976)]}\BibitemShut {NoStop}%
\bibitem [{Note1()}]{Note1}%
  \BibitemOpen
  \bibinfo {note} {If one chooses nonzero $a$, discontinuity of $S(b)$ at $b=0$
  and $b=b^*$ leads to two $\delta $-function contributions of opposite signs
  in the angle $\varphi $, which cancel each other in the transverse
  cross-section $\sigma _\perp $.}\BibitemShut {Stop}%
\bibitem [{\citenamefont {Blonder}\ \emph {et~al.}(1982)\citenamefont
  {Blonder}, \citenamefont {Tinkham},\ and\ \citenamefont {Klapwijk}}]{BTK}%
  \BibitemOpen
  \bibfield  {author} {\bibinfo {author} {\bibfnamefont {G.~E.}\ \bibnamefont
  {Blonder}}, \bibinfo {author} {\bibfnamefont {M.}~\bibnamefont {Tinkham}}, \
  and\ \bibinfo {author} {\bibfnamefont {T.~M.}\ \bibnamefont {Klapwijk}},\
  }\href@noop {} {\bibfield  {journal} {\bibinfo  {journal} {Phys. Rev. B}\
  }\textbf {\bibinfo {volume} {25}},\ \bibinfo {pages} {4515} (\bibinfo {year}
  {1982})}\BibitemShut {NoStop}%
\bibitem [{\citenamefont {Kopnin}\ and\ \citenamefont
  {Kravtsov}(1976{\natexlab{b}})}]{Kop76}%
  \BibitemOpen
  \bibfield  {author} {\bibinfo {author} {\bibfnamefont {N.~B.}\ \bibnamefont
  {Kopnin}}\ and\ \bibinfo {author} {\bibfnamefont {V.~E.}\ \bibnamefont
  {Kravtsov}},\ }\href@noop {} {\bibfield  {journal} {\bibinfo  {journal} {Zh.
  Eksp. Teor. Fiz.}\ }\textbf {\bibinfo {volume} {71}},\ \bibinfo {pages}
  {1664} (\bibinfo {year} {1976}{\natexlab{b}})},\ \bibinfo {note} {[{Sov.
  Phys.--JETP}, {\bf 44}, 861 (1976)].}\BibitemShut {Stop}%
\bibitem [{\citenamefont {Gal'perin}\ and\ \citenamefont {Sonin}(1976)}]{Gal}%
  \BibitemOpen
  \bibfield  {author} {\bibinfo {author} {\bibfnamefont {Y.~M.}\ \bibnamefont
  {Gal'perin}}\ and\ \bibinfo {author} {\bibfnamefont {E.~B.}\ \bibnamefont
  {Sonin}},\ }\href@noop {} {\bibfield  {journal} {\bibinfo  {journal} {Fiz.
  Tverd. Tela (Leningrad)}\ }\textbf {\bibinfo {volume} {18}},\ \bibinfo
  {pages} {3034} (\bibinfo {year} {1976})},\ \bibinfo {note} {[{Sov.
  Phys.--Solid State} {\bf 18}, 1768 (1976)]}\BibitemShut {NoStop}%
\bibitem [{\citenamefont {Fisher}\ and\ \citenamefont {Pickett}(2009)}]{FP}%
  \BibitemOpen
  \bibfield  {author} {\bibinfo {author} {\bibfnamefont {S.~N.}\ \bibnamefont
  {Fisher}}\ and\ \bibinfo {author} {\bibfnamefont {G.~R.}\ \bibnamefont
  {Pickett}},\ }in\ \href@noop {} {\emph {\bibinfo {booktitle} {Progress in Low
  Temperature Physics: Quantum Turbulence}}},\ \bibinfo {series} {Progress in
  Low Temperature Physics}, Vol.~\bibinfo {volume} {16},\ \bibinfo {editor}
  {edited by\ \bibinfo {editor} {\bibfnamefont {B.}~\bibnamefont {Halperin}}\
  and\ \bibinfo {editor} {\bibfnamefont {M.}~\bibnamefont {Tsubota}}}\
  (\bibinfo  {publisher} {Elsevier},\ \bibinfo {year} {2009})\ pp.\ \bibinfo
  {pages} {147--194}\BibitemShut {NoStop}%
\bibitem [{\citenamefont {Suramlishvili}\ \emph {et~al.}(2012)\citenamefont
  {Suramlishvili}, \citenamefont {Baggaley}, \citenamefont {Barenghi},\ and\
  \citenamefont {Sergeev}}]{BarAnd}%
  \BibitemOpen
  \bibfield  {author} {\bibinfo {author} {\bibfnamefont {N.}~\bibnamefont
  {Suramlishvili}}, \bibinfo {author} {\bibfnamefont {A.~W.}\ \bibnamefont
  {Baggaley}}, \bibinfo {author} {\bibfnamefont {C.~F.}\ \bibnamefont
  {Barenghi}}, \ and\ \bibinfo {author} {\bibfnamefont {Y.~A.}\ \bibnamefont
  {Sergeev}},\ }\href {\doibase 10.1103/PhysRevB.85.174526} {\bibfield
  {journal} {\bibinfo  {journal} {Phys. Rev. B}\ }\textbf {\bibinfo {volume}
  {85}},\ \bibinfo {pages} {174526} (\bibinfo {year} {2012})}\BibitemShut
  {NoStop}%
\bibitem [{\citenamefont {Sonin}(1997)}]{PRB7}%
  \BibitemOpen
  \bibfield  {author} {\bibinfo {author} {\bibfnamefont {E.~B.}\ \bibnamefont
  {Sonin}},\ }\href@noop {} {\bibfield  {journal} {\bibinfo  {journal} {Phys.
  Rev. B}\ }\textbf {\bibinfo {volume} {55}},\ \bibinfo {pages} {485} (\bibinfo
  {year} {1997})}\BibitemShut {NoStop}%
\bibitem [{\citenamefont {Landau}\ and\ \citenamefont {Lifshitz}(1982)}]{LLqu}%
  \BibitemOpen
  \bibfield  {author} {\bibinfo {author} {\bibfnamefont {L.~D.}\ \bibnamefont
  {Landau}}\ and\ \bibinfo {author} {\bibfnamefont {E.~M.}\ \bibnamefont
  {Lifshitz}},\ }\href@noop {} {\emph {\bibinfo {title} {Quantum mechanics}}}\
  (\bibinfo  {publisher} {Pergamon Press},\ \bibinfo {year} {1982})\BibitemShut
  {NoStop}%
\bibitem [{\citenamefont {Kulik}(1969)}]{Kul}%
  \BibitemOpen
  \bibfield  {author} {\bibinfo {author} {\bibfnamefont {I.~O.}\ \bibnamefont
  {Kulik}},\ }\href@noop {} {\bibfield  {journal} {\bibinfo  {journal} {Zh.
  Eksp. Teor. Fiz.}\ }\textbf {\bibinfo {volume} {57}},\ \bibinfo {pages}
  {1745} (\bibinfo {year} {1969})},\ \bibinfo {note} {[Sov. Phys.--JETP {\bf
  30}, 944 (1970]}\BibitemShut {NoStop}%
\bibitem [{\citenamefont {Ishii}(1970)}]{Ishi}%
  \BibitemOpen
  \bibfield  {author} {\bibinfo {author} {\bibfnamefont {C.}~\bibnamefont
  {Ishii}},\ }\href@noop {} {\bibfield  {journal} {\bibinfo  {journal} {Prog.
  Theor. Phys.}\ }\textbf {\bibinfo {volume} {44}},\ \bibinfo {pages} {1525}
  (\bibinfo {year} {1970})}\BibitemShut {NoStop}%
\bibitem [{\citenamefont {Bardeen}\ and\ \citenamefont {Johnson}(1972)}]{Bard}%
  \BibitemOpen
  \bibfield  {author} {\bibinfo {author} {\bibfnamefont {J.}~\bibnamefont
  {Bardeen}}\ and\ \bibinfo {author} {\bibfnamefont {J.~L.}\ \bibnamefont
  {Johnson}},\ }\href@noop {} {\bibfield  {journal} {\bibinfo  {journal} {Phys.
  Rev B}\ }\textbf {\bibinfo {volume} {5}},\ \bibinfo {pages} {72} (\bibinfo
  {year} {1972})}\BibitemShut {NoStop}%
\bibitem [{\citenamefont {Lamb}(1997)}]{Lam}%
  \BibitemOpen
  \bibfield  {author} {\bibinfo {author} {\bibfnamefont {H.}~\bibnamefont
  {Lamb}},\ }\href@noop {} {\emph {\bibinfo {title} {Hydrodynamics}}}\
  (\bibinfo  {publisher} {Cambridge University Press},\ \bibinfo {year}
  {1997})\BibitemShut {NoStop}%
\bibitem [{\citenamefont {Volovik}(1998)}]{Vol98}%
  \BibitemOpen
  \bibfield  {author} {\bibinfo {author} {\bibfnamefont {G.~E.}\ \bibnamefont
  {Volovik}},\ }\href@noop {} {\bibfield  {journal} {\bibinfo  {journal}
  {Pis'ma Zh. Eksp. Teor. Fiz.}\ }\textbf {\bibinfo {volume} {67}},\ \bibinfo
  {pages} {502} (\bibinfo {year} {1998})},\ \bibinfo {note} {[{JETP Lett.},
  {\bf 67}, 528 (1998)]}\BibitemShut {NoStop}%
\bibitem [{\citenamefont {Aronov}\ \emph {et~al.}(1981)\citenamefont {Aronov},
  \citenamefont {Gal'perin}, \citenamefont {Gurevich},\ and\ \citenamefont
  {Kozub}}]{Aro}%
  \BibitemOpen
  \bibfield  {author} {\bibinfo {author} {\bibfnamefont {A.~G.}\ \bibnamefont
  {Aronov}}, \bibinfo {author} {\bibfnamefont {Y.~M.}\ \bibnamefont
  {Gal'perin}}, \bibinfo {author} {\bibfnamefont {V.~L.}\ \bibnamefont
  {Gurevich}}, \ and\ \bibinfo {author} {\bibfnamefont {V.~I.}\ \bibnamefont
  {Kozub}},\ }\href@noop {} {\bibfield  {journal} {\bibinfo  {journal} {Adv.
  Phys.}\ }\textbf {\bibinfo {volume} {30}},\ \bibinfo {pages} {539} (\bibinfo
  {year} {1981})}\BibitemShut {NoStop}%
\bibitem [{\citenamefont {Golubchik}\ \emph {et~al.}(2012)\citenamefont
  {Golubchik}, \citenamefont {Polturak},\ and\ \citenamefont {Koren}}]{Pol}%
  \BibitemOpen
  \bibfield  {author} {\bibinfo {author} {\bibfnamefont {D.}~\bibnamefont
  {Golubchik}}, \bibinfo {author} {\bibfnamefont {E.}~\bibnamefont {Polturak}},
  \ and\ \bibinfo {author} {\bibfnamefont {G.}~\bibnamefont {Koren}},\ }\href
  {\doibase 10.1103/PhysRevB.85.060504} {\bibfield  {journal} {\bibinfo
  {journal} {Phys. Rev. B}\ }\textbf {\bibinfo {volume} {85}},\ \bibinfo
  {pages} {060504} (\bibinfo {year} {2012})}\BibitemShut {NoStop}%
\end{thebibliography}
%

\end{document}